\documentclass[journal]{IEEEtran}
\usepackage{graphicx}
\input{epsf}

\newcommand{\ga}{ \lower .75ex \hbox{$\sim$} \llap{\raise .27ex \hbox{$>$}} }
\newcommand{\la}{ \lower .75ex \hbox{$\sim$} \llap{\raise .27ex \hbox{$<$}} }

\def\beq{\begin{equation}}
\def\eeq{\end{equation}}

\begin{document}

\title{The impact of polarized extragalactic radio sources on the
  detection of CMB anisotropies in polarization}

\author{Marco~Tucci$^a$~and~Luigi~Toffolatti$^{b,c}$
\thanks{$^a$~LAL, Univ Paris--Sud, CNRS/IN2P3, Orsay, France; 
tucci@lal.in2p3.fr}%
\thanks{$^b$~Departamento de
  F\'\i{sica}, Universidad de Oviedo, c. Calvo Sotelo s/n, 33007
  Oviedo, Spain; ltoffolatti@uniovi.es}
\thanks{$^c$~I.F.C.A. --
Universidad de Cantabria, avda. los Castros s/n - 39005 Santander,
Spain}}

\maketitle

\begin{abstract}
Recent polarimetric surveys of extragalactic radio sources (ERS) at
frequencies $\nu\ga1$\,GHz are reviewed. By exploiting all the most
relevant data on the polarized emission of ERS we study the frequency
dependence of polarization properties of ERS between 1.4 and
86\,GHz. For flat--spectrum sources the median (mean) fractional
polarization increases from 1.5\% (2--2.5\%) at 1.4\,GHz to 2.5-3\%
(3--3.5\%) at $\nu>10$ GHz. Steep--spectrum sources are typically more
polarized, especially at high frequencies where Faraday depolarization
is less relevant. As a general result, we do not find that the
fractional polarization of ERS depends on the total flux density at
high radio frequencies, i.e $\geq 20$ GHz. Moreover, in this frequency
range, current data suggest a moderate increase of the fractional
polarization of ERS with frequency. A formalism to estimate ERS
number counts in polarization and the contribution of unresolved
polarized ERS to angular power spectra at Cosmic Microwave Background
(CMB) frequencies is also developed and discussed. As a first
application, we present original predictions for the {\it Planck}
satellite mission. Our current results show that only a dozen
polarized ERS will be detected by the {\it Planck} Low Frequency
Instrument (LFI), and a few tens by the High Frequency Instrument
(HFI). As for CMB power spectra, ERS should not be a strong
contaminant to the CMB E--mode polarization at frequencies
$\nu\ga70\,$GHz. On the contrary, they can become a relevant
constraint for the detection of the cosmological B--mode polarization
if the tensor-to-scalar ratio is $\la0.01$.
\end{abstract}

\section{Introduction}
\label{s1}

The radiation emitted by extragalactic radio sources (ERS) at high
radio frequencies, mainly synchrotron radiation from
relativistic electrons in their jets and lobes, can be highly
polarized, with an intrinsic degree of linear polarization as high
as $\sim 70-75\%$ in homogeneous sources with an unidirectional
magnetic field (see, e.g., \cite{ginz65,ginz69}, for comprehensive
discussions on the subject).

These have to be considered as maximum values, which could be
detected only in the case of very homogeneous sources with highly
aligned magnetic field lines, which are unlikely to be observed
among ERS. Actually, a much lower degree of total linear
polarization, $P$, is commonly observed in ERS at cm or mm
wavelengths (see e.g., \cite{tay07,mur10}), with only very few ERS
showing a total fractional polarization, $\Pi=P/S$, as high as
$\sim$10\% of the total flux density, $S$.

Nevertheless, this low fractional polarization observed in ERS may
constitute a problem for the detection of primordial
polarization in the Cosmic Microwave Background (CMB), since ERS are
the dominant polarized foreground at small angular scales and the
CMB polarized signal is only a few percent of CMB temperature
anisotropy. Therefore, CMB polarization studies need a careful
determination of the polarized emission from foreground sources in
general and of ERS in particular.

CMB polarized anisotropies can be decomposed into modes of even-parity
(E-mode) and odd-parity (B-mode).
As for the E-modes, generated by scalar perturbations in the
primordial universe, they were first detected by \cite{kov02} and
all the following observations (e.g., \cite{page07,dunk09}) have
confirmed measurements of CMB E-modes compatible with the
concordance $\Lambda$CDM model. The more elusive B-modes are
generated by tensor metric perturbations, i.e. ``primordial''
gravitational waves generated during inflation, and, according to
predictions of inflationary models, with an amplitude directly
proportional to the energy scale at which the inflation occurred
\cite{zal97,kam97}. A detection of these ``primordial'' B-modes
would provide the first real measure of the energy scale of
inflation: thus, it would produce a real breakthrough in modern
cosmology. But the detection of the B-modes CMB polarization --
parameterized by the tensor-to-scalar ratio, $T/S=r$ -- is still a
great challenge. In fact, the {\it Planck} mission
\cite{tau10,Planck11a}\footnote{Detailed discussions of the {\it
    Planck} Low Frequency Instrument (LFI) and High Frequency
  Instrument (HFI) expected polarization capabilities and calibrations
  are given by \cite{lea10,ros10}.} would be marginally able to detect
tensor metric perturbations by the direct detection of the primordial
CMB B-mode, but only in the case of very high values of the
tensor-to-scalar ratio of primordial perturbations, $r\ga 0.05-0.1$
\cite{efst09}. Otherwise, we will have to rely on proposed future
space experiments, like $COrE$ \cite{CORE}, specifically designed to
detect CMB polarization by virtue of a much higher sensitivity than
current ones.

The above discussion clearly illustrates the importance of giving
the best up-to-date estimate of the average/median fractional polarization
as well as of the distributions of fractional polarization
\cite{tuc04}, $\Pi_i$, at least for each one of the different ERS
populations (\cite{tof98,dez05,tuc11}) which are providing relevant
contributions to temperature anisotropies of the CMB.

Statistical studies of polarized emission from ERS have shown that at
$\sim$1.4-5 GHz, the fractional polarization, $\Pi$, increases with
decreasing flux density
\cite{tay07,tuc04,mes02,sad06,sub10}. Different explanations have been
proposed of this result: e.g., a population change at fainter flux
density \cite{mes02,sub10} or a changing fraction of radio-quiet AGN
\cite{tay07}. However, the cause of this increase in $\Pi$ is still
unknown. On the other hand, it has been shown by \cite{tuc04} that
this increase of the degree of polarization with decreasing flux
density is only observed, at cm wavelengths, for steep--spectrum ERS,
i.e sources with $\alpha < -0.5$ if $S(\nu)\propto \nu^{\alpha}$, and
not for flat--spectrum ERS ($\alpha\geq -0.5$).

Other very recent studies of the polarized emission in ERS have tried
to analyze the dependence of the fractional polarization with
luminosity, redshift and the source environment. The current, still
preliminary, results show no correlation between the fractional
polarization and redshift, whereas a weak correlation is found between
decreasing luminosity and increasing degree of polarization
\cite{ban11}. According to the analysis of highly polarized elliptical
galaxies \cite{shi10} no differences have been found in the source
environments between low polarization and ultrahigh polarization
sources. This result should indicate that a high polarization must be
a result of intrinsic properties of ERS.

The outline of the paper is as follows: in Sect.\,\ref{s2} we briefly
summarize the main processes giving rise to linear and circular
polarization in ERS; in Sect.\,\ref{s3} we present current published
data on the polarized emission from ERS; Sect.\,\ref{s4} is dedicated
to discuss our current results on statistical properties of the
polarized emission in ERS; in Sect.\,\ref{s6} we give a short review
on the most recent cosmological evolution model for ERS;
Sect.\,\ref{s7} presents our predictions on the contributions of ERS
polarized radiation, given by unresolved sources, to the E-- and B--
modes; finally, in Sect.\,\ref{s8}, we present our conclusions.

\section{Polarized emission from ERS}
\label{s2}

Observations as well as statistical studies of the fractional
polarization of ERS are very interesting on their own, and not only
because ERS constitute a major contaminant of the CMB
polarized signal. In fact, they are an important tool in active
galactic nuclei (AGN) research, as they give valuable information on the
physical quantities which determine the characteristics of the
synchrotron radiation emitted by relativistic electrons accelerated
by homogeneous and/or random magnetic fields in AGN
\cite{ginz65,ginz69}.

On the one hand, precise measurements of the total synchrotron radiation
emitted by a radio source give an estimate - under some assumption,
i.e. equipartition between the field and particle energy - of the
total magnetic field strength. On the other hand, the degree of
polarization of the radio wave provides information on the direction
of the main magnetic field in the source environment and also on
its degree of ordering. The magnetic field structure can, in turn,
provide information on the relationship between the environments of
the ERS and their properties.

As already noted, compact ERS typically show a degree of total
linear polarization of a few percent of their radio total intensity.
Therefore, magnetic fields in radio sources are believed to be
highly inhomogeneous, or almost without ordering, although the
observed non-vanishing linear polarization gives an indication of a
certain degree of ordering of the field \cite{ruz02,gard66}.

The precise orientation of magnetic fields lines inside jets and lobes
is still unknown, but theoretical arguments as well as observational
evidence show that magnetic fields are indeed partially ordered. As
for observations, these conclusions are based on measurements of the
orientations of the linear polarization, revealing coherent structures
across the images. From the theoretical point of view, an ordered
magnetic field is expected when shocks compress an initially random
field (with the field $\mathbf{B}$ perpendicular to the jet axis) or
when such initial fields are sheared to lie in a plane, with $\mathbf{
  B}$ parallel to the jet axis \cite{hog79,lai80,lai81,mar85}.

Circular polarization (CP), which is a common feature of quasars and
BL\,Lac objects (or simply BL\,Lacs), commonly known as
blazars\footnote{Blazar sources are jet-dominated extragalactic
  objects -- observed within a small angle of the jet axis -- in which
  the beamed component dominates the observed emission \cite{ang80}.},
is preferentially generated near synchrotron self-absorbed jet cores
and is detected in about 30\%--50\% of these sources
\cite{beck02,ruz02}. In these inhomogeneous, optically thick,
synchrotron sources, the emission of the electron population at lower
energies is hidden by self-absorption. These invisible electrons
produce {\it Faraday rotation} and conversion, which is the most
likely mechanism capable of creating the observed CP in blazar sources
\cite{beck02,gard66}. In any case, the measured degrees of CP are
generally well below the levels of linear polarization \cite{ray00},
and thus negligible, at least at GHz frequencies.


The change in the position angle of the linearly polarized radiation
which passes through a magneto-ionic medium, i.e. the {\it Faraday
  rotation}, can be expressed by the rotation angle
$\Delta\phi$[rad]=$RM$[rad/m$^2$]$\,\lambda^2$[m$^2$]
\cite{gard66,strom73} experienced by the polarization vector, where
$RM$ indicates the {\it rotation measure}. The rotation measure, $RM$,
is the line-of-sight integral $RM$[rad/m$^2$]$=0.81\int
n_e$[cm$^{-3}$]$B_{\|}$[$\mu$G]$d\ell$[pc], $n_e$ being the electron
density and $B_{\|}$ the component of the magnetic field along the
line-of-sight \cite{bur06}.

If the rotation depth is the same for all the emission volumes of the
radio source, the net result is a rotation equal to $RM\,\lambda^2$ of
the direction of polarization, without any effect on the degree of
polarization. To change the degree of polarization there must be a
variation in depth, either along or transverse to the line-of-sight
\cite{gard66}.

Faraday rotation is commonly observed towards extragalactic radio
sources. It was first observed by \cite{coop62}, who found an RM of
$-60$ rad/m$^2$ towards the center of Centaurus A. More recently,
observations of the rest-frame RM in AGN cores and jets reported
values from a few hundreds to several thousands rad/m$^2$ (e.g.,
\cite{zav03,zav04,bat11}). On the other hand, ERS dominated by
emission from radio lobes show lower values of RMs, i.e.. from a few
tens to hundreds of rad/m$^2$ (e.g., \cite{tay09}).


\section{Polarization data on ERS at cm and mm wavelenghts}
\label{s3}

\subsection{Polarization surveys at 1.4\,GHz}

A large--scale polarization catalogue of radio sources is provided by
the NRAO VLA Sky Survey (NVSS) at 1.4\,GHz \cite{con98}. This survey
covers $\Omega\simeq10.3$\,sr of the sky with
$\delta\ge-40^{\circ}$. The catalogue contains the flux density $S$
and the Stokes parameters $Q$ and $U$ of almost $2\times10^6$ discrete
sources with $S\ga 2.5$\,mJy. Extensive analyses of these data were
carried out by \cite{mes02} and \cite{tuc04}.  By correlating NVSS
sources with sources from the Green Bank 4.85\,GHz (GB6) catalogue
\cite{gre96}, statistical polarization properties have been derived
for a sub--sample of $\sim 30,000$ ERS with $S_{1.4GHz}\ge100$\,mJy,
divided into steep-- and flat--spectrum sources. Steep--spectrum
sources are found to increase the degree of polarization ($\Pi$) with
decreasing flux density: in fact, the median (mean) $\Pi$ value
increases from $\Pi\simeq1.1\%$ ($\simeq2\%$) for $S>800$\,mJy up to
$\Pi\simeq1.8\%$ ($\simeq2.7\%$) for the faintest ERS of the sample,
at $100\le S<200$\,mJy. On the other hand, flat--spectrum sources show
a lower fractional polarization (median $\sim1.3\%$ and mean
$\simeq2\%$) with no significant trend with flux density \cite{tuc04}.

Two recent sub--mJy surveys at 1.4\,GHz with polarization
measurements, i.e. the Dominion Radio Astrophysical Observatory (DRAO)
{\it Planck} Deep Fields project \cite{tay07,gra10,ban11} and the
Australia Telescope Low--Brightness Survey (ATLBS) \cite{sub10}, seem
to confirm the above result of a higher level of fractional
polarization in steep--spectrum sources with fainter flux
densities. However, there is not a clear explanation for the origin of
the anticorrelation between $\Pi$ and flux density in this source
population.

\begin{figure}
\centering
\includegraphics[width=9cm]{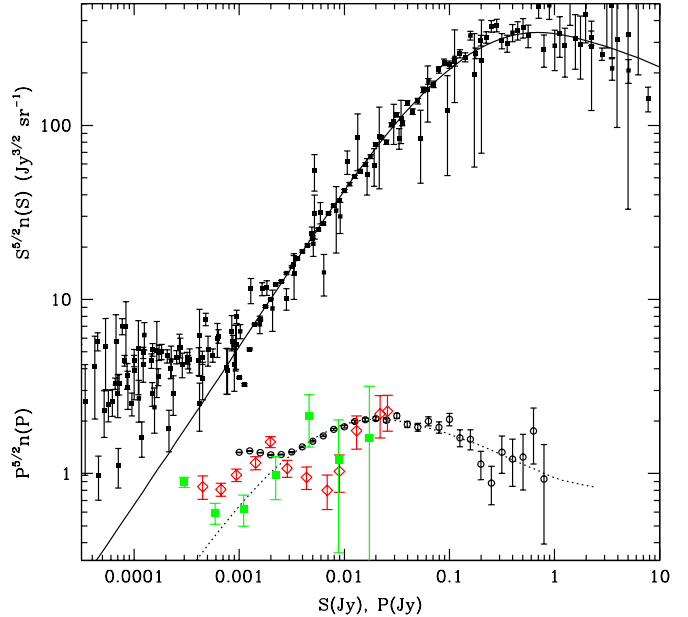}
\caption{Differential number counts at 1.4\,GHz as a function of total
  intensity (upper points) and of polarized intensity (lower points:
  black empty circles from NVSS; red empty diamonds from DRAO {\it
    Planck} Deep Fields; green squares from ATLBS). The continuous curve
  represents the number counts of AGNs from \cite{mas10}, whereas the dotted
  curve is obtained by the previous counts and assuming a constant
  fractional polarization of 3.3\% for all ERS.}
\label{f1}
\end{figure}

In Fig.\,\ref{f1} we plot differential number counts at 1.4\,GHz for
the polarized intensity, computed from the surveys presented
above. The continuous curve represents a fit to total number counts of
AGNs from \cite{mas10}, whereas the dotted curve is obtained from the
previous fit by assuming a constant fractional polarization of
$\Pi=3.3$\% for {\it all} ERS. This value is chosen in order to fit
NVSS data; the fit is extremely good down to polarized intensities of
$P\simeq3\,$mJy. At fainter $P$ fluxes the predicted curve bends down
whereas observational data keep a flatter shape, in agreement with a
higher fractional polarization for ERS at fainter flux density levels.

\subsection{Multiwavelength samples of polarized ERS}

\subsubsection{Linear polarization in a sample of the B3--VLA survey}

Polarization measurements of sources in the B3--VLA survey
\cite{vig89} were carried out by \cite{kle03} at 10.5, 4.85 and
2.7\,GHz using the Effelsberg 100--m telescope. Taking into account
only sources with $S_{10.5GHz}\ge80$\,mJy, a sample of 106 objects
(out of 208) were defined with detected polarization. They found that
flat--spectrum sources are significantly less polarized than
steep--spectrum ones at 10.5\,GHz. The latter ERS population shows a
median fractional polarization that strongly depends on the frequency,
from $\sim2\%$ at 1.4\,GHz to $\sim6\%$ at 10.5\,GHz, indicating a
high Faraday depolarization at lower frequencies. Moreover, they
notice that compact steep--spectrum sources exhibit much stronger
depolarization than non--compact ones, and that sources showing larger
linear size are more polarized. On the other hand, flat--spectrum
sources are characterized by almost constant and low degrees of
polarization ($\sim2.5\%$) over the whole wavelength range considered.

\subsubsection{The Australia Telescope 20\,GHz (AT20G) survey}

The AT20G survey is a blind survey of the whole Southern sky at
20\,GHz with follow--up observations at 4.8 and 8.6\,GHz carried out
with the Australia Telescope Compact Array (ATCA) \cite{mur10}. The
full source catalogue includes 5890 sources detected above a
flux--density limit of 40\,mJy. The AT20G catalogue is found to be
91\% complete above 100\,mJy and 79\% above 50\,mJy in regions south
of declination $-15^{\circ}$. Polarization was detected at 20\,GHz for
768 sources, 467 of which also have simultaneous polarization
detections at 5 and/or 8\,GHz (upper limits on the polarized flux
density are provided for non--detections, as described in
\cite{mur10}). Taking into account upper limits, a median (mean)
fractional polarization of 2.6\% (2.7\%) at 20\,GHz is found for
sources with $S_{20GHz}>250$\,mJy, whereas an average 17\%
depolarization is observed at 5\,GHz with respect to 20\,GHz
\cite{mas11}. Compact sources with detected polarization are separated
into flat-- and steep--spectrum sources: the mean values of $\Pi$ are
2.9\% and 3.8\% respectively, thus confirming the higher fractional
polarization in steep--spectrum sources.

\subsubsection{VLA polarization measurements of WMAP point sources}

Using the VLA, \cite{jac10} carried out polarization measurements at
8.4, 22 and 43\,GHz of a complete sample of ERS brighter than 1\,Jy in
the 5--year WMAP catalogue and with declinations north of
$-34^{\circ}$. The sample consists of 203 objects: polarized emission
was detected for 123, 169 and 167 sources at 8.4, 22 and 43\,GHz
respectively (at 8.4\,GHz only a subset of 134 were observed) and 105
were detected at all the 3 frequencies. An accurate analysis of the
statistical properties of the polarized intensity of the sample is
done by \cite{bat11}, including an analysis of the correlations
between the fractional polarization and the spectral indices. Here
below, we summarize the main results obtained by them:

\begin{itemize}

\item The distribution of the fractional polarization varies slightly
  as a function of the frequency. Including  sources undetected in
  polarization, the mean fractional polarization is
  $\langle\Pi\rangle=2.9$ 3.0 and 3.5 percent at 8.4, 22 and 43\,GHz.

\item No correlation is found between the fractional polarization at
  22\,GHz and the intensity spectral indices, $\alpha_8^{22}$ and
  $\alpha_{22}^{43}$ (We remind readers that the sample, selected at 22\,GHz,
  is dominated by flat--spectrum sources).

\item There is a significant change in the polarization angle
  between 8.4 and 22\,GHz. For 45 sources, the position angle satisfies
  the $\lambda^2$ dependence, and intrinsic rotation measures in
  excess of 1000\,rad\,m$^{-2}$ were observed for a large number of
  them.

\item Polarization of 71 sources were also measured at 86\,GHz by
  \cite{agu10}. The fractional polarization is typically higher at
  86\,GHz than at 43\,GHz.

\end{itemize}

\subsubsection{VLA polarization measurements in an ACT survey field}

In \cite{saj11}, VLA observations in total intensity and polarization
at 4.86, 8.46, 22.46 and 43.34\,GHz are presented. The sample is
selected from the AT20G survey and consists of 159 ERS, out of the
almost 200 sources with flux density $S_{20GHz}>40$\,mJy in a field of
the Atacama Comsmology Telescope (ACT) survey. Polarized emission for
about 60 sources was detected at all 4 frequencies, whereas the
detections are 141, 146, 89 and 59 from low to high
frequencies. Fractional polarization distributions are very similar at
5 and 8.5\,GHz, whereas a trend of increasing polarization with
increasing frequency is indicated at 22 and 43\,GHz. In particular, a
tail of strongly polarized ($\ga10\%$) sources is observed at
43\,GHz. Data at 22 and 43\,GHz suggest also that the polarization
fraction in steep--spectrum sources is significant higher than in
flat--spectrum sources.

\subsection{Other samples of polarized ERS at high frequencies}

\begin{itemize}

\item \cite{ric04} presented polarization observations of 250 (out of
  258) southern sources in the complete 5--GHz 1--Jy sample of
  \cite{kuh81} by using the ATCA facilities at 18.5\,GHz. Polarized
  flux densities were measured for 170 sources (114 flat--spectrum and
  56 steep--spectrum), upper limits were set for an additional subset
  of 27 sources (12 flat--spectrum and 15 steep--spectrum), and 53
  sources were rejected (probably extended objects). The final
  flat--spectrum sample is almost complete (80\%), while only 49\% of
  steep--spectrum sources have reliable detections. In the
  flat--spectrum sample the median fractional polarization is
  $\simeq2.7\%$ and the mean $\simeq2.9\%$, and no sources have
  $\Pi>10\%$. The median fractional polarization for the
  flat--spectrum sources included in the NVSS catalogue is about a
  factor 2 lower than that at 18.5\,GHz. However, a relevant increase
  in the polarization fraction at 18.5\,GHz is noticed only for sources
  with $\Pi\la1\%$ at 1.4\,GHz.

\item \cite{agu10} presented a 3.5\,mm polarimetric survey of ERS
  using the IRAM 30\,m Telescope. Their sample consists of 145
  flat--spectrum sources with $\delta>-30^{\circ}$ and flux density
  $\ga1\,$Jy at 86\,GHz. Linear polarization is detected for 76\% of
  the sample (110 objects). They found that BL\,Lacs
  ($\Pi_{median}\simeq4.4\%$) are more strongly polarized than quasars
  ($\Pi_{median}\simeq3.1\%$). This result seems to be in
  contradiction with the idea that quasars should be more polarized at
  high frequencies than BL\,Lacs because in the latter sources the
  synchrotron self--absorbed spectrum is mantained up to higher
  frequencies. A possible explaination provided by the authors comes
  from the recent evidence that the view angle of jets in quasars is
  smaller than that in BL\,Lacs \cite{hov09,pus09}. So, if the
  magnetic field is not homogeneous along the jet, a lower fractional
  polarization level is expected from sources better oriented to the
  line of sight (i.e., quasars).

  Moreover, for those sources with detected polarization at 15\,GHz,
  they found that $\Pi_{86\,GHz}$ is larger than $\Pi_{15\,GHz}$ by a
  median factor $\approx2$, and about 20\% of sources have the
  $\Pi_{86\,GHz}/\Pi_{15\,GHz}$ ratio larger than 4. \cite{agu10}
  suggest that this increase may be explained by a combination of two
  phenomena: the 86--GHz emission comes from a region with greater
  degree of order in the magnetic field; or/and, at 15\,GHz, emission
  is still affected by Faraday depolarization.

\item \cite{lop09} looked for polarized sources in the {\it WMAP}
  five--year data, using a new technique named {\it filtered fusion}.
  They detected polarization in 13 ERS at a confindence level
  $\ge99\%$ and polarized flux density higher than 300\,mJy.

\end{itemize}

\section{Statistical properties of the polarized emission in ERS}
\label{s4}

In order to provide reliable estimates of ERS contamination to CMB
anisotropy polarization measurements, we need to address the following
main questions about polarization properties of ERS: 1) how the
fractional polarization, $\Pi$, varies from cm to mm wavelengths; 2)
{\it if} the fractional polarization depends on the flux density; 3)
how polarization properties change among the different populations of
ERS. In this Section we address these questions on the basis of the
observational data presented and discussed in the previous Section.

\subsection{Flat--spectrum sources}
\label{s41}

\begin{table*}
\caption{Median, mean and $<\Pi^2>^{1/2}$ of the fractional
  polarization for flat--spectrum sources in the almost
  complete sub--sample of the AT20G surveys at different
  observational frequencies and flux ranges. $N_{tot}$ refers to the
  total number of objects in the corresponding flux range; $N_{multi}$
  to the number of objects with 5-- and 8--GHz measurements; $N_{fl}$
  to the number of flat--spectrum sources; $N_{det}$ to the number
  of flat--spectrum sources with polarization detection.}
\centering
\begin{tabular}{cccccccccc}
\hline
$S$(Jy) & & $N_{tot}$ & $N_{multi}$ & & $N_{fl}$ & $N_{det}$ &
$\Pi_{med}$ & $<\Pi>$ & $<\Pi^2>^{1/2}$ \\
\hline
\hline
 & & \multicolumn{8}{c}{20\,GHz} \\
\hline
$\ge1.0$ & & 130 & 114 & & 110 & 85 & 2.05 & 2.82 & 3.72 \\
$\ge0.5$ & & 315 & 287 & & 264 & 188 & 2.01 & 2.76 & 3.84 \\
$[$0.5,\,1) & & 185 & 173 & & 154 & 103 & 1.92 & 2.72 & 3.92 \\
\hline
 & & \multicolumn{8}{c}{8.6\,GHz} \\
\hline
$\ge1.0$ & & & &  & 110 & 87 & 2.00 & 2.52 & 3.00 \\
$\ge0.5$ & & & &  & 264 & 180 & 1.76 & 2.34 & 2.85 \\
$[$0.5,\,1) & & & &  & 154 & 93 & 1.54 & 2.21 & 2.73 \\
\hline
 & & \multicolumn{8}{c}{4.8\,GHz} \\
\hline
$\ge1.0$ & & & &  & 110 & 93 & 1.90 & 2.31 & 2.71 \\
$\ge0.5$ & & & &  & 264 & 186 & 1.71 & 2.25 & 2.69 \\
$[$0.5,\,1) & & & &  & 154 & 93 & 1.59 & 2.20 & 2.68 \\
\hline
\end{tabular}
\label{t1}
\end{table*}

We investigate more deeply the polarization of ERS observed in the
AT20G survey. We consider the almost--complete sample at
$\delta<-15^{\circ}$ and, when available, we use 5-- and 8--GHz
measurements to separate ERS into steep-- and flat--spectrum
sources. In Table\,\ref{t1} we report statistical properties of the
fractional polarization (i.e., the mean $<\Pi>$, the median
$\Pi_{med}$ and $<\Pi^2>^{1/2}$) for flat--spectrum sources as a
function of the flux density range and the frequency. They are
computed using the Survival Analysis techniques and the Kaplan--Meyer
estimator as implemented in the ASURV code \cite{lav92}, which takes
into account upper limits on the fractional polarization for
estimating the above quoted quantities. In fact, when polarization is
not detected, an upper limit is provided. About 10\% of sources in the
AT20G sample we use here have measurements only at 20\,GHz (see
Table\,\ref{t1}). Although most of them should be flat--spectrum
sources, for a more consistent comparison with results at 5 and 8\,GHz
we prefer to exclude them from the analysis.

At 20\,GHz the median and the mean fractional polarization do not
present any significant variation between the subsamples defined by
$S\ge1$\,Jy and $0.5\le S<1$\,Jy. At the lower frequencies, $<\Pi>$
and $\Pi_{med}$ show a moderate decrease at fainter flux densities. In
particular, two--sample tests implemented in the ASURV code (e.g., the
Gehan's generalized Wilcoxon test and the Peto\,\&\,Prentice
generalized Wilcoxon test) yield a probability of $\sim10\%$ and
$\sim1\%$ that the distributions of the fractional polarization for
$S\ge1$\,Jy and $0.5\le S<1$\,Jy at 5 and 8\,GHz respectively are
drawn from the same parent distribution (compared to a probability of
30\% at 20\,GHz). At flux densities lower than 500\,mJy, the high
number of upper limits ($\ga50$\%) makes our estimates unreliable.

\begin{figure*}
\centering
\includegraphics[width=6.5cm]{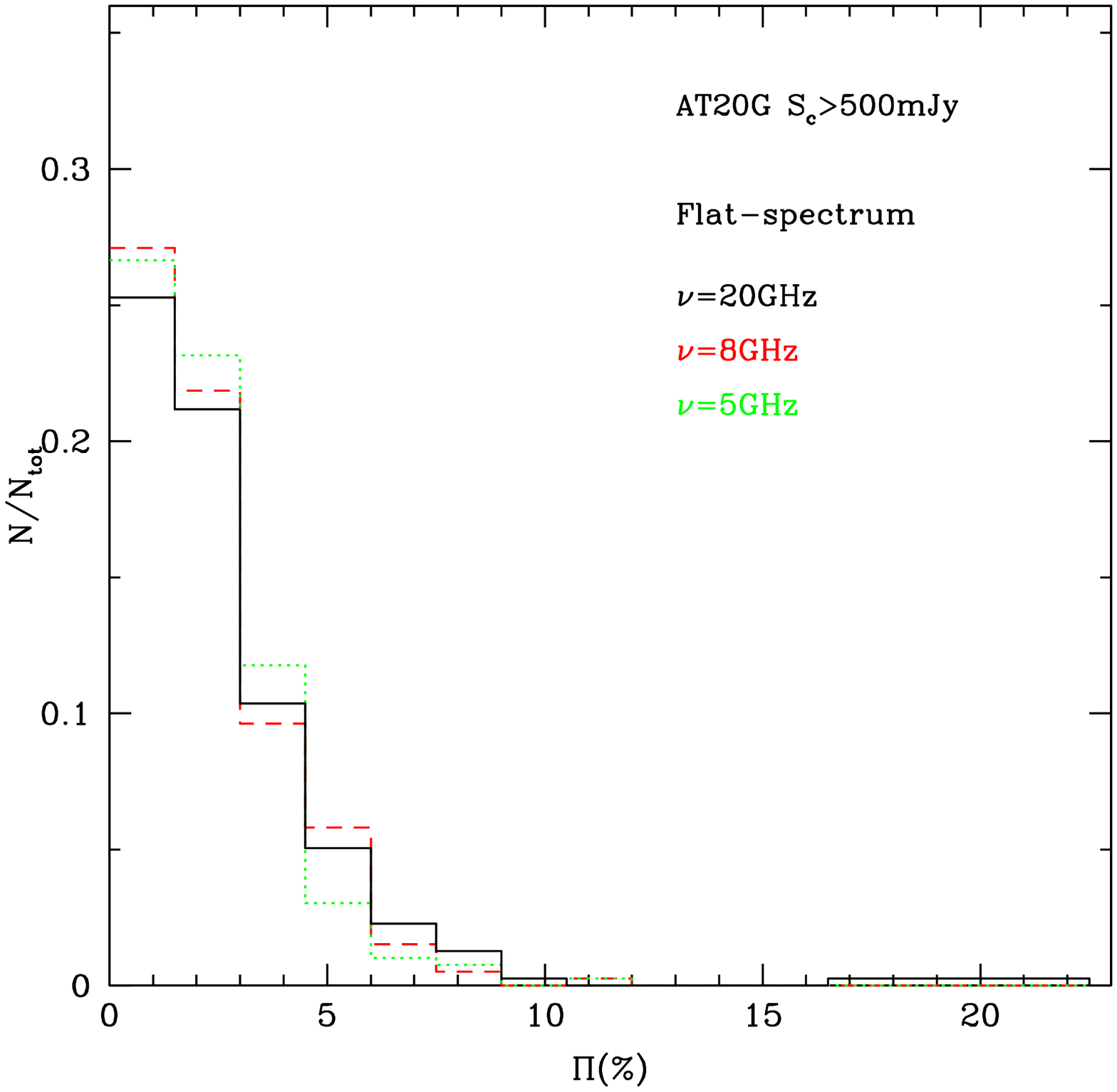}
\includegraphics[width=6.5cm]{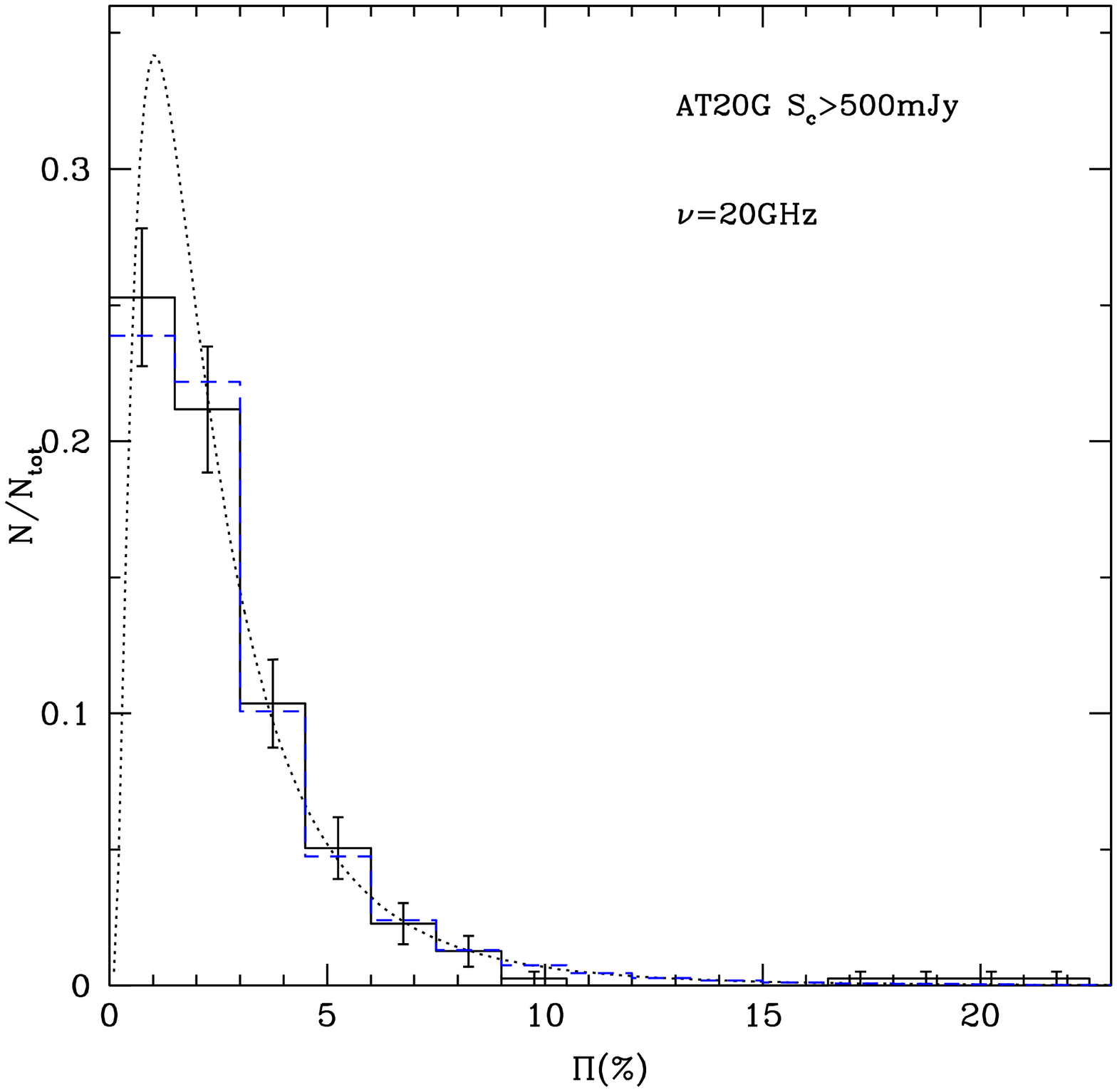}
\caption{{\it Left panel}: distribution of the fractional polarization
  at 4.8, 8.6 and 20\,GHz (green dotted, red dashed and black solid
  histograms, respectively) for flat--spectrum sources with
  $S\ge500\,$mJy in the almost--complete sub--sample of the AT20G
  survey. {\it Right panel}: distribution of the fractional
  polarization at 20\,GHz (black solid histogram) compared to the
  distribution (blue dashed histogram) produced by a log--normal
  distribution (see also Section\,\ref{s61}) with $\Pi_{med}=2.01$ and
  $<\Pi^2>=3.84$ (black dotted line; see Table\,\ref{t2} at 20\,GHz.}
\label{f2a}
\end{figure*}

On the other hand, a larger fractional polarization is observed as the
frequency increases, with, on average, $\approx$18\% of depolarization
at 5\,GHz with respect to 20\,GHz, in agreement with results from
\cite{mas11}. The ASURV two--sample tests yield
only a 10\% probability to have the same parent distribution for $\Pi$
at 5 and 20\,GHz. For a better comparison, we show in
Fig.\,\ref{f2a} the distributions of the fractional polarization
discussed above as obtained by using the Kaplan--Meyer
estimator. These distributions are well fitted by a log--normal
distribution with $\Pi_{med}$ and $<\Pi^2>$ values taken from
Table\,\ref{t1} (see, e.g., the case at 20\,GHz in
Fig.\,\ref{f2a}).\footnote{Fig.\,\ref{f2a} clearly shows that the
  log-normal distribution, when averaged in each $\Pi$ bin, is giving
  predictions always compatible with the observed $\Pi$ values well
  inside the 1$\sigma$ level, except for very few bins at high $\Pi$
  levels where the statistics is very poor. We have also verified that
  adopting other distribution functions, e.g. a truncated gaussian, we
  are not able to reproduce equally well the observed distributions of
  $\Pi$.}

\begin{figure*}
\centering
\includegraphics[width=6.5cm]{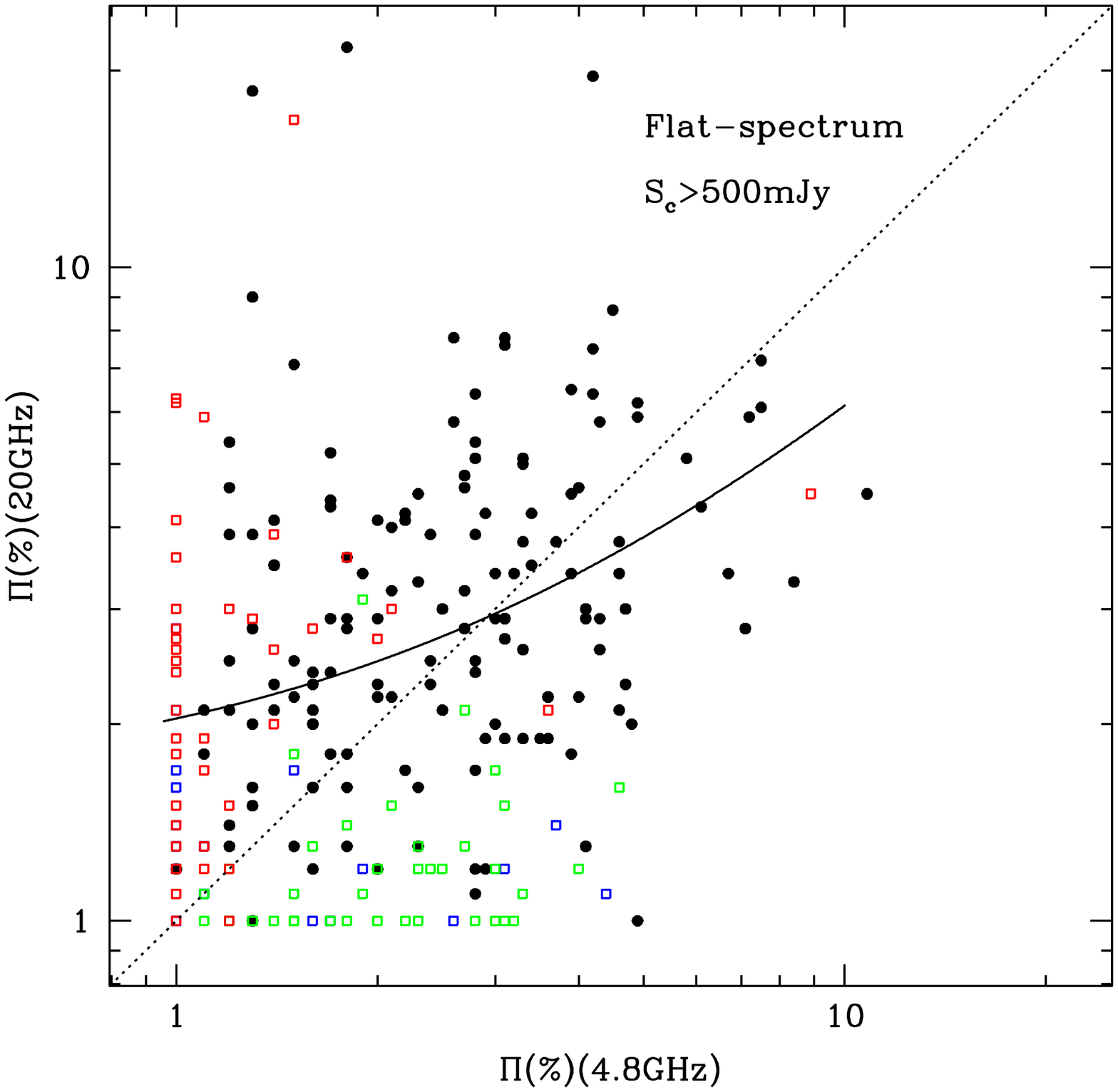}
\includegraphics[width=6.5cm]{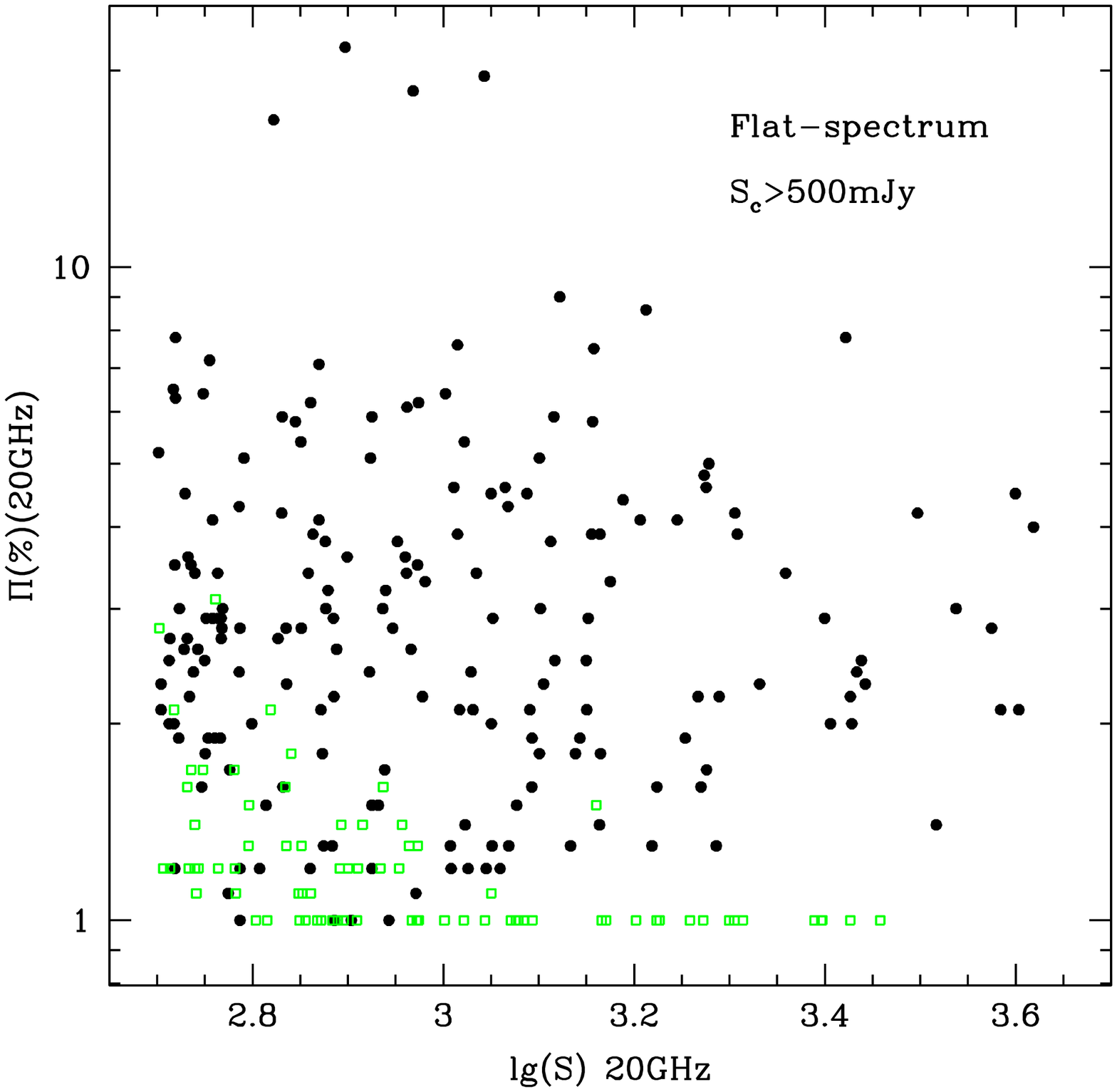}
\caption{{\it Left panel}: correlation between fractional
  polarizations at 4.8 and 20\,GHz for flat--spectrum sources with
  $S\ge500\,$mJy. Green points indicate upper limits in $\Pi$ at
  20\,GHz, red points at 4.8\,GHz, and blue points at the both
  frequencies. The solid line is the linear regressions found by the
  ASURV code (see the text). {\it Right panel}: the fractional
  polarization at 20\,GHz as a function of the flux density at the
  same frequency (green points indicate upper limits in $\Pi$).}
\label{f2}
\end{figure*}

In Fig.\,\ref{f2} we also study the correlation of $\Pi_{20GHz}$
with $\Pi_{4.8GHz}$ and $S_{20GHz}$: no clear correlation is found
between the fractional polarization and the flux density at 20\,GHz
(the generalized Kendall's tau test yields a probability $P\sim60\%$
of no correlation). On the other hand, as expected, we see a strong
correlation between the fractional polarization at 20 and 4.8\,GHz
(with a $P<0.01\%$ of no correlation).  Using the Schmitt's method
from the ASURV code, we find a linear regression
$\Pi_{20GHz}=1.58+0.46\Pi_{4.8GHz}$. This result seems to indicate
that only sources with very low fractional polarization at 5\,GHz
(i.e., $\Pi_{4.8GHz}\la2\%$) have a significant increase of $\Pi$ at
20\,GHz (see also Fig.\,\ref{f2}). The large offset term in the
linear regression we find, however, could be also partially due to
Eddington bias in the AT20G catalogue and to the large number of
sources with upper limits in fractional polarization at $\Pi\ga1$\%.

In Fig.\,\ref{f4} we plot the mean and median values of the
fractional polarization as obtained from the surveys presented in the
previous section. These data allow us to cover frequencies from 1.4 to
86\,GHz. Although there is large scatter in the data, we see a general
increasing trend for the median fractional polarization: $\Pi_{med}$
is $\sim1.5\%$ at 1.4\,GHz, around 2--2.5\% in the range 5--10\,GHz,
and finally 2--$3\%$ at $\nu>10\,GHz$. The mean fractional
polarization has a more linear increase with the frequency in all the
samples, with $<\Pi>$ varying from 2--2.5\% at 1.4\,GHz to 3--$3.5\%$
at $\nu\ge20$\,GHz.

Most of the data presented in Fig.\,\ref{f4} are coming from samples
of bright sources, typically with $S\ga1\,$Jy. Samples with fainter
sources are the B3--VLA ($S_c=80\,$mJy) and the sample discussed in
\cite{saj11}. The latter one provides polarization for sources with
$S\ge40\,$mJy. However, since at the frequencies 20 and 43\,GHz the
number of polarized ERS detected in the sample is less than
50\,percent, we have decided to consider only flat--spectrum sources
with $S\ge80$\,mJy. Spectral indices are estimated using flux
densities at 4.86 and 8.46\,GHz. Table\,\ref{t2} reports the number of
detections and the corresponding values of the mean, median and
$<\Pi^2>^{1/2}$ of the fractional polarization. For the two highest
frequencies of the sample we estimate the median in two ways: firstly,
we use the ASURV code, by taking into account the upper limits;
secondly, we assume a fractional polarization $\leq 1$\% for those
sources without measured polarization (values indicated in brackets in
Table\,\ref{t2}). The large spread between the two values at 43\,GHz
is an indication that the sample, also including upper limits, could
provide biased values of statistical properties of $\Pi$ at these
frequencies.

If we compare results from the previous two samples (i.e., B3--VLA and
\cite{saj11}) with surveys of bright sources, we cannot find any clear
evidence of higher fractional polarization in faint sources. However,
larger and deeper samples of data are clearly required to settle the
question.

In Fig.\,\ref{f4} we also include the $<\Pi^2>^{1/2}$ values. This
quantity is important in order to estimate the angular power spectra
of the polarized signal due to undetected ERS. At $\nu>20$\,GHz,
i.e. the most interesting frequencies for CMB data analyses, sources
have $<\Pi^2>^{1/2}\sim4\%$. In \cite{agu10} optical identifications
are provided for ERS in the \cite{saj11} sample. Over a total of 145
objects, 107 are identified as quasars and 26 as BL\,Lacs: we find
$\Pi_{med}=3.0\%$, 3.6\% and $<\Pi^2>^{1/2}=3.8\%$, 4.5\% for FSRQs
and BL\,Lacs, respectively.

\begin{table}
\caption{Median and mean of the fractional polarization and
  $<\Pi^2>^{1/2}$ from the Sajina et al. (2011) sample. For
  flat--spectrum sources we consider only objects with
  $S\ge80$\,mJy. Median values in brackets are computed by assigning a
  value of $\Pi<1\%$ to sources without a detection in polarization.}
\centering
\begin{tabular}{ccccccc}
\hline
\multicolumn{7}{c}{Flat--spectrum sources} \\
\hline
$\nu$[GHz] & $N_{sou}$ & $N_{det}$ & \multicolumn{2}{c}{$\Pi_{med}$} &
$<\Pi>$ & $<\Pi^2>^{1/2}$ \\
\hline
4.86 & 56 & 54 & 2.16 & & 2.47 & 2.97 \\
8.46 & 56 & 55 & 2.25 & & 2.61 & 3.10 \\
22.46 & 56 & 44 & 3.12 & (2.88) & 3.79 & 4.21 \\
43.34 & 50 & 33 & 4.14 & (2.64) & 4.85 & 5.79 \\
\hline
\multicolumn{7}{c}{Steep--spectrum sources} \\
\hline
4.86 & 45 & 44 & 2.56 & & 4.10 & 5.79 \\
8.46 & 45 & 44 & 3.48 & & 4.67 & 5.97 \\
22.46 & 44 & 25 & 6.14 & (3.40) & 7.17 & 8.30 \\
43.34 & 28 & 11 & 6.77 & & 7.18 & 8.03 \\
\hline
\end{tabular}
\label{t2}
\end{table}

\subsection{Steep--spectrum sources}

\begin{figure*}
\centering
\includegraphics[width=6.5cm]{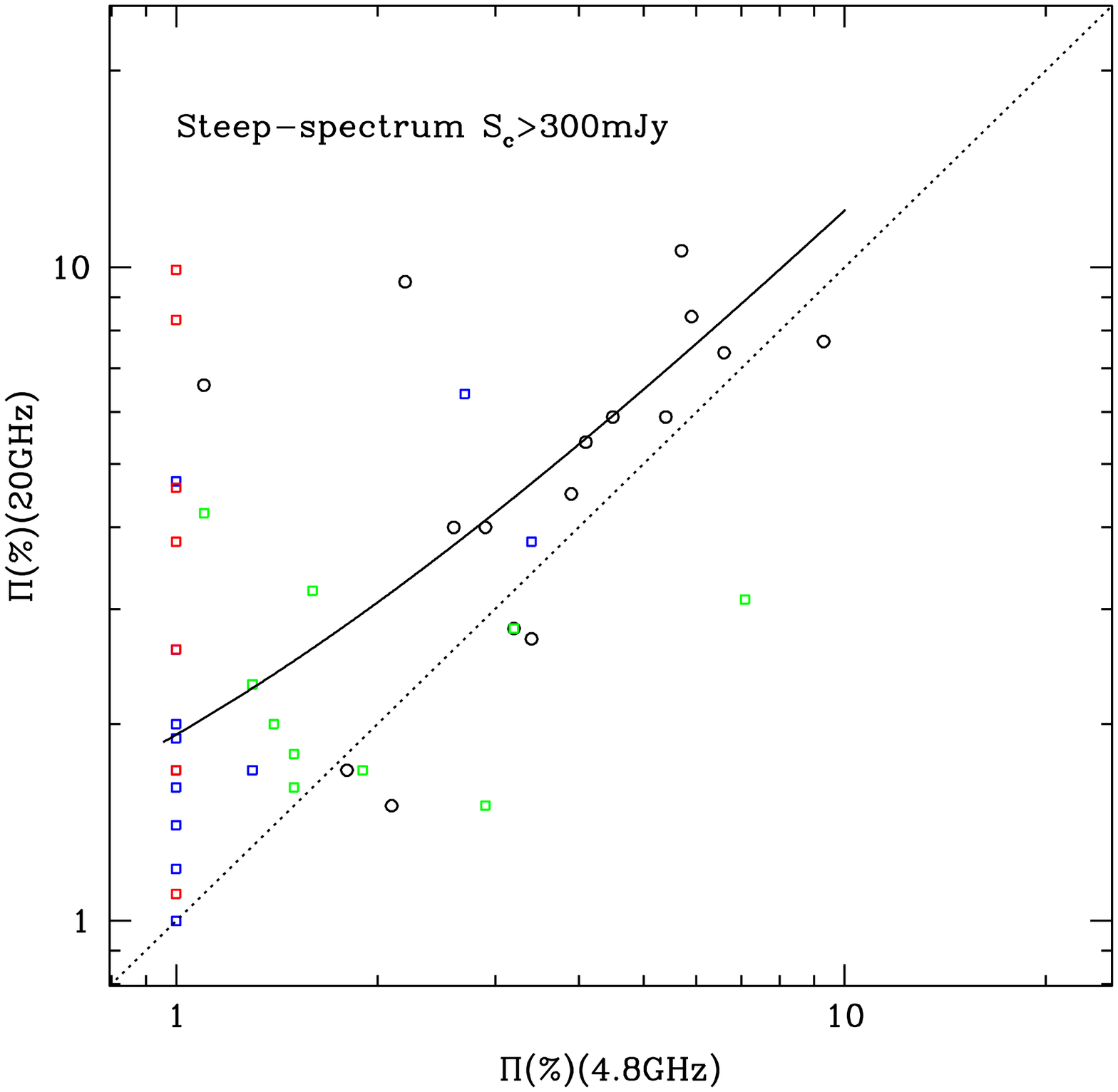}
\includegraphics[width=6.5cm]{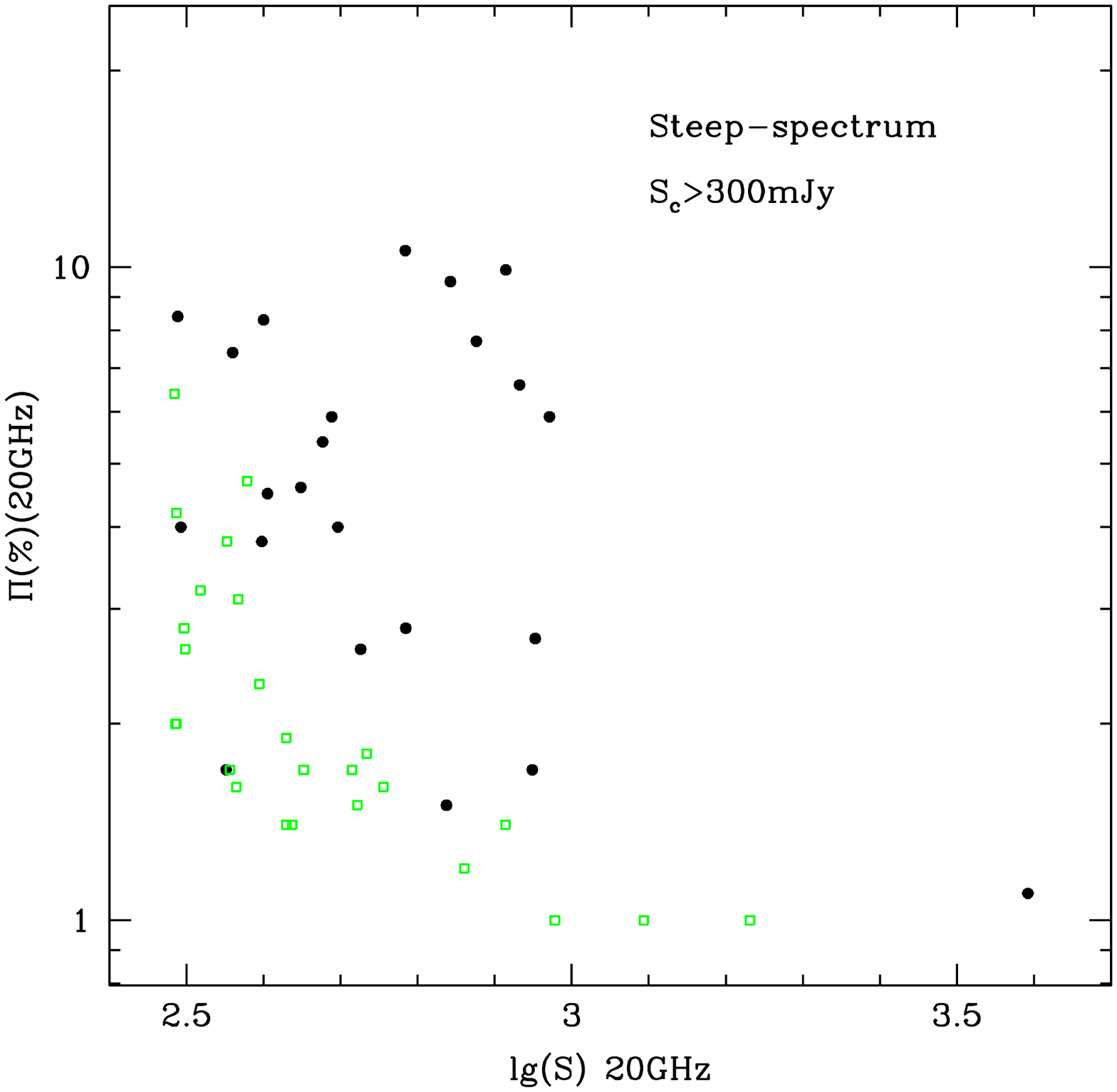}
\caption{Fractional polarization for steep--spectrum sources at
  20\,GHz as a function of fractional polarization at 4.8\,GHz (left
  panel) and of flux density at 20\,GHz (right panel). Points of
  different colours have the same meaning as in Fig.\,\ref{f2}.}
\label{f3}
\end{figure*}

\begin{figure*}
\centering
\includegraphics[width=9.cm]{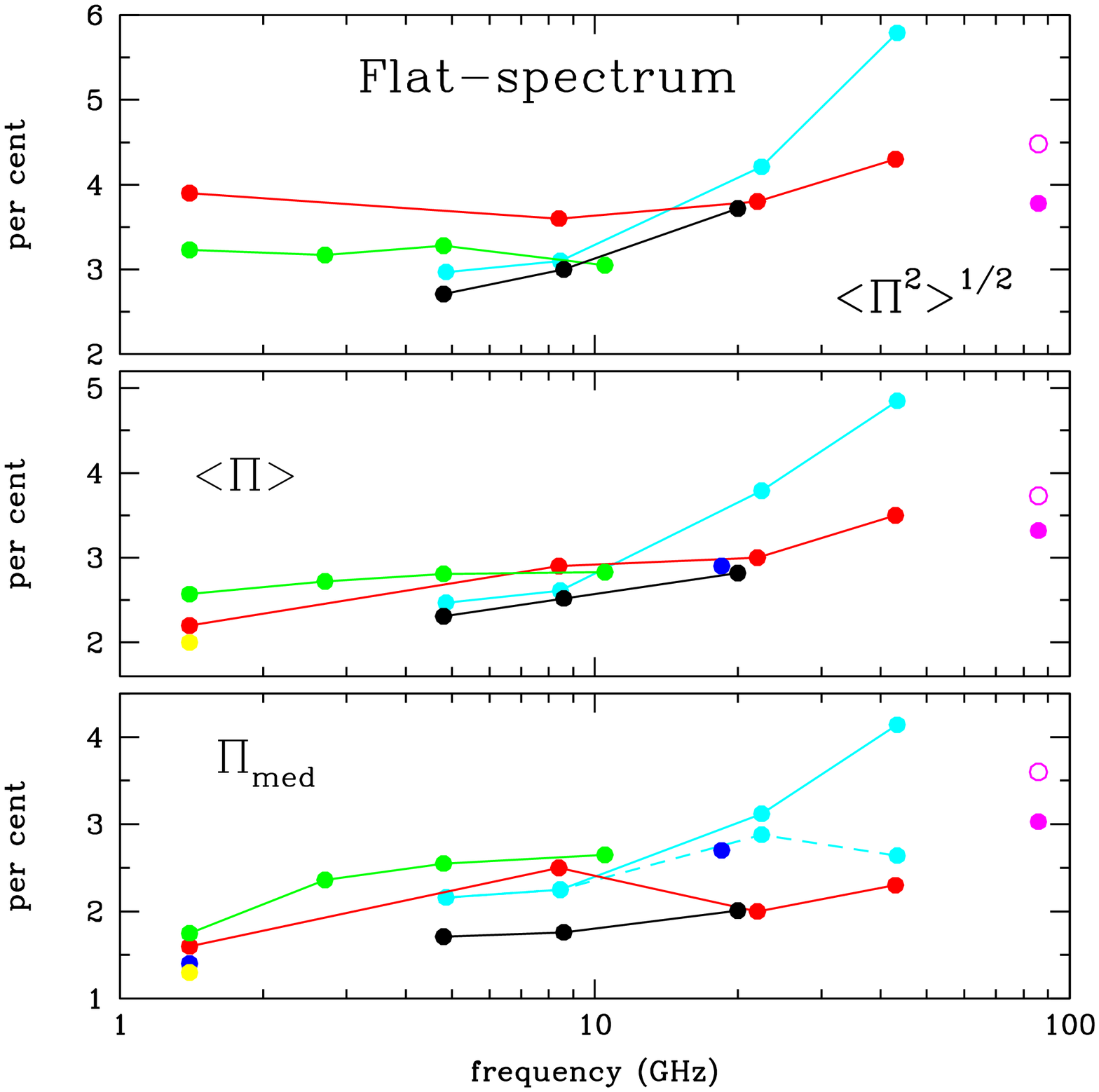}
\includegraphics[width=9.cm]{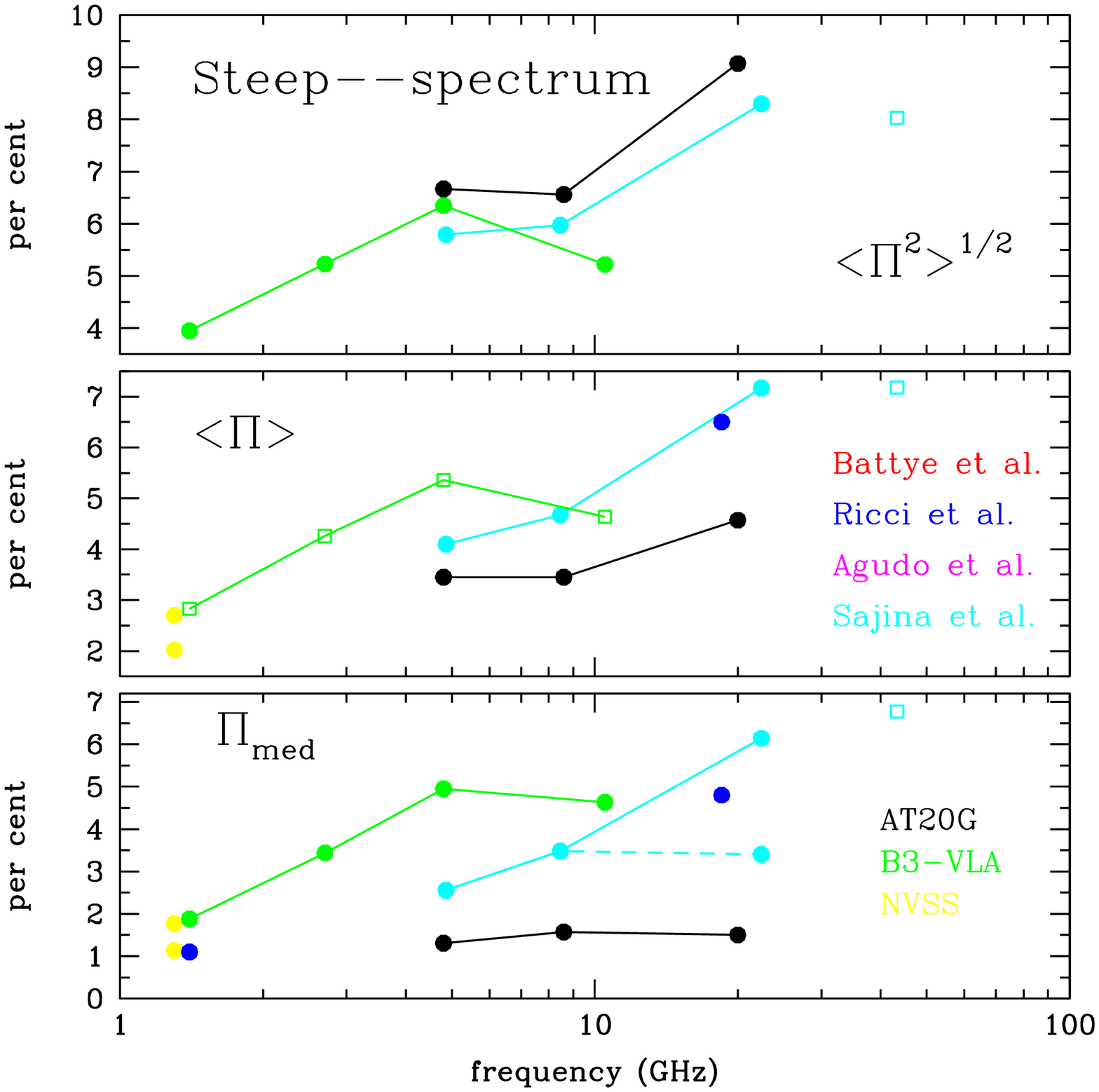}
\caption{Median, mean and $<\Pi^2>^{1/2}$ of fractional polarization
  for flat-- (upper panels) and steep--spectrum (lower panels)
  sources. {\it Black points} refer to the almost complete sub--sample
  of the AT20G survey with $S_c=500$\,mJy for flat--spectrum sources
  and 300\,mJy for steep--spectrum sources. {\it Red points} refer to
  data presented in \cite{bat11}. {\it Green points} refer to the
  B3--VLA sample (respect to \cite{kle03}, we distinguish flat-- and
  steep--sources on the basis of their spectral index between 1.4 and
  5\,GHz; values shown in the plot are obtained using the ASURV code,
  taking into account upper limits in the fractional
  polarization). {\it Cyan points} refer to the sample from
  \cite{saj11} (see text and Table\,\ref{t2}; empty cyan points at
  43\,GHz for the steep--spectrum have to be considered as upper
  limits). {\it Blue points} refer to results from \cite{ric04} and
  {\it magenta points} from \cite{agu10} (solid and empty points are
  for FSRQs and BL\,Lacs, respectively). {\it Yellow points} refer to
  the NVSS survey (see \cite{tuc04}): for steep--spectrum sources we
  plot results for sources with $100\le S<200\,$mJy and
  $S\ge800$\,mJy.}
\label{f4}
\end{figure*}

The number of steep--spectrum sources with polarization measurements
becomes very small at frequencies $\ga10\,$GHz, preventing any study
of the correlation between fractional polarization and flux density
(see, e.g., Fig.\,\ref{f3}). In the sub--sample of AT20G at
$\delta<-15^{\circ}$ there are 51 steep--spectrum sources with
$S\ge0.3\,$Jy and only 25 of them have polarization detected at
20\,GHz. We find that the median fractional polarization for these
sources is less than 2\% at all the frequencies (see Fig.\,\ref{f4},
lower panel). These very small values could be biased due to the small
number of detected sources and to the incompleteness of the sample at
faint flux densities. Nevertheless, from Fig.\,\ref{f3} we can see a
general increase of the fractional polarization between 4.8 and
20\,GHz: the ratio of $\Pi$ at 20 and 4.8\,GHz is typically close to
one ($<1.5$) for sources with $\Pi_{4.8GHz}>2\%$, but become higher
for sources weakly polarized at 4.8\,GHz. In fact, the linear
regression found by the Schmitt's method implemented in the ASURV code
yields $\Pi_{20GHz}=0.78+1.14\Pi_{4.8GHz}$.

From Fig.\,\ref{f4} we can observe the strong increase of the
fractional polarization from 1.4\,GHz to 5\,GHz, where Faraday
depolarization is probably very relevant. The increase becomes more
moderate up to 20\,GHz.  The large difference in the median values
among different samples is perhaps related to the small samples
considered and the large fraction of sources with upper limits in
polarization. The largest sample of steep--spectrum sources is
provided by B3--VLA (77 sources, all with detected polarization) and
the Ricci et al. sample (71 sources, 15 of them with upper limits): in
these samples the median fractional polarization is $\sim5\%$ between
5--20\,GHz and the mean varies from 5 to 6.5\%.

\subsection{Number counts in polarization of ERS at 20\,GHz}
\label{s5}

We provide a first estimate of number counts in polarization, $n(P)$,
at 20\,GHz by exploiting WMAP and AT20G polarization source
catalogues. For very bright sources, we use the polarized source
sample detected by \cite{lop09} in the 5-yr WMAP CMB anisotropy maps
(see their Table\,2). We exclude from number counts Fornax\,A,
Virgo\,A and Centaurus\,A because they are all local objects. Then, we
consider the nearly--complete subsample of AT20G at
$\delta<-15^{\circ}$ and $S\ge50$\,mJy. This sample allows us to
compute $n(P)$ down to polarized fluxes of $\sim10$\,mJy. Number
counts, that are reported in Table\,\ref{t3} (see also
Fig.\,\ref{f5}), have been corrected for the estimated
incompleteness of the sample (i.e., the completeness is 0.91 for
$S\ge100\,$mJy and 0.79 for $50\le S<100\,$mJy \cite{mur10}). As
displayed by Fig.\,\ref{f5}, number counts in total polarization, $n(P)$,
are almost flat at 20 GHz, at least in the flux density range in which
they can be estimated.

Given that an upper limit on the fractional polarization $\Pi_{up}$ is
always provided for the ERS in this sample (see \cite{mur10}, for more
details), except for 5 very bright objects with no information on
polarization, we also give a tentative estimate (Table\,\ref{t3},
third column) of their contribution to the number counts $n_i^{up}(P)$
between $P_i$ and $P_i+\Delta P$ by means of
\begin{equation}
n_i^{up}(P)=\sum_k\int_{\Pi_i}^{\Pi_f} {\mathcal P}(\Pi)\,d\Pi\,,
\label{ee0}
\end{equation}
with $\Pi_i=P_i/S_k$ and $\Pi_f={P_i+\Delta P \over S_k}$.
The sum is done over all the sources without polarization detection
and $S_k$ is the flux density of the k--th object. ${\mathcal P}(\Pi)$
is the probability function for the fractional polarization: we take a
log--normal function with $\Pi_{med}=2.0$\% and $<\Pi^2>^{1/2}=2.8$\%
(see Table\,\ref{t1} and Fig.\,\ref{f2}) if $\Pi\le\Pi_{up}$ and
zero otherwise. For the 5 objects without any polarization
information, no upper limits is considered.

As shown by Table\,\ref{t3} and Fig.\,\ref{f5}, the contribution of
sources undetected in polarization (displayed by empty squares) is
generally negligible, except for bins at very faint polarization
levels (i.e. less than 20\,mJy), and for bins at $P\ga100\,$mJy, due
to the 5 sources without polarization information and flux density
between 1 and 10\,Jy. Only for the faintest bin $n_i^{up}$ is larger
than the uncertainty on the number counts. Therefore, we can conclude
that number counts in polarization estimated from the AT20G sample are
not affected by upper limits in the polarized flux density.

\begin{table}
\caption{Number counts in polarization at 20\,GHz from WMAP and AT20G
  source catalogues. $N_{det}$ indicates the number of objects with
  detected polarization, while $n^{up}(P)$ is the estimated
  contribution of AT20G objects with no polarization information or
  with upper limits in the fractional polarization.}
\centering
\begin{tabular}{cccccc}
\hline
\multicolumn{6}{c}{AT20G sub--sample} \\
\hline
$P$(mJy) & & $N_{det}$ & $n(P)$ & $\sigma_{n(P)}$ & $n^{up}(P)$ \\
\hline
   11.2 & &  87 & 0.828E+04 & 0.889E+03 & 0.205E+04 \\
   14.1 & &  81 & 0.607E+04 & 0.675E+03 & 0.636E+03 \\
   17.8 & &  57 & 0.338E+04 & 0.447E+03 & 0.263E+03 \\
   22.4 & &  49 & 0.231E+04 & 0.329E+03 & 0.102E+03 \\
   28.2 & &  33 & 0.123E+04 & 0.215E+03 & 0.242E+02 \\
   35.5 & &  23 & 0.683E+03 & 0.142E+03 & 0.112E+02 \\
   44.7 & &  22 & 0.519E+03 & 0.111E+03 & 0.104E+02 \\
   56.2 & &  18 & 0.337E+03 & 0.795E+02 & 0.938E+01 \\
   70.8 & &  17 & 0.253E+03 & 0.613E+02 & 0.842E+01 \\
   89.1 & &  10 & 0.118E+03 & 0.374E+02 & 0.752E+01 \\
  112.2 & &   5 & 0.469E+02 & 0.210E+02 & 0.664E+01 \\
  141.3 & &   3 & 0.224E+02 & 0.129E+02 & 0.385E+01 \\
  177.8 & &   4 & 0.237E+02 & 0.118E+02 & 0.184E+01 \\
  223.9 & &   2 & 0.941E+01 & 0.665E+01 & 0.121E+01 \\
  281.8 & &   1 & 0.374E+01 & 0.370E+01 & 0.755E+00 \\
\hline
\multicolumn{6}{c}{5--yr WMAP sample} \\
\hline
398.1 & & 5 & 0.199E+01 & 0.889E+00 & \\
631.0 & & 2 & 0.530E+00 & 0.375E+00 & \\
1072. & & 1 & 0.150E+00 & 0.150E+00 & \\
\hline
\end{tabular}
\label{t3}
\end{table}

\section{Cosmological evolution models and high--frequency number
  counts of ERS}
\label{s6}

Early evolutionary models of radio sources
\cite{dan87,dun90,tof98,jac99} were able to give remarkably
successful fits to the majority of data coming from surveys at
$\nu\la10$\,GHz, and down to flux densities of a few mJy. More
recently, \cite{dez05} and \cite{mas10} exploited the wealth of
newly available data on luminosity functions, multi-frequency source
counts and redshift distributions to provide new cosmological
evolution models of radio sources at frequencies $\ga5$\,GHz and
$\la5\,$GHz, respectively. These two models are based on the
determination of the epoch--dependent luminosity functions for
different source populations, starting from the local luminosity
function and by adopting different luminosity evolution laws with
free parameters fitted from observational data.

The predictions of high--frequency number counts of ERS provided by
the above evolution models assume a simple power--law spectrum for
ERS. Each source population is characterized by an ``average'',
fixed, spectral index, or by two spectral indices (at most). This
``classical'' modeling has to be considered as a first -- although
successful -- approximation, but it gives rise to an increasing
mismatch with observed high--frequency ($> 30$\,GHz) number counts
currently available.

Indeed, the ERS energy spectra can be quite different from a single
power--law if analyzed in large frequency intervals. Different
mechanisms can be responsible for this: a) a spectral steepening due
to the more rapid energy loss of high--energy electrons with source
age, i.e. ``electron ageing''; b) a transition from the optically
thick to the optically thin regime at high radio frequencies; c)
at different wavelengths radio emission can be dominated by
different components characterized by distinct spectral behaviors
\cite{tuc08}. In particular, a clear steepening at mm wavelengths is
theoretically expected for radio flat spectra of AGN core emission
\cite{kel66,bla79}. This steepening has been already observed in
blazars (see, e.g., \cite{Planck11k}) and has also been
statistically suggested by recent analyses of different ERS samples
at $\nu>30$\,GHz \cite{wal07,gon08,mas10c,Planck11i,tuc11}.

A first attempt of taking such steepening in blazar spectra into
account has been done by \cite{tuc11}. In this work the spectral
behaviour of blazars at mm wavelengths is statistically described by
considering the main physical mechanisms responsible for the
emission. In agreement with classical models of the synchrotron
emission in the inner jets of blazars these authors interprete the
high--frequency steepening observed in blazar spectra as caused, at
least partially, by the transition from the optically--thick to the
optically--thin regime in the observed spectra at mm wavelengths.
Based on the published models of synchrotron emission from
inhomogeneous, unresolved, relativistic jets
\cite{bla79,kon81,mar85}, the value of the frequency $\nu_M$ at
which the spectral break occurs is estimated as a function of the
relevant physical parameters of AGNs: the redshift, the Doppler
factor, and the linear dimension of the region (approximated as
homogeneous and spherical) that is mainly responsible for the
emission at the break frequency.

Recent high frequency data \cite{vie10,mar11}, and in particular the
first 1.6 {\it Planck} survey \cite{Planck11i} have provided new
important results on number counts and related statistics of ERS, and
they can be used to constrain the different possible cases/models
featuring a spectral break in the emission of AGN jets (see
\cite{tuc11}). The most successful one (indicated in \cite{tuc11} as
C2Ex, i.e. the relevant emission is emitted from a more "extended"
region in the inner jet of FSRQs) assumes different distributions of
the break frequency for BL\,Lacs and FSRQs. According to this model,
most of the FSRQs should show a bend in their otherwise flat spectra
between 10 and 100\,GHz, whereas in BL\,Lac spectral breaks should be
typically observed at $\nu>100\,$GHz (implying that the observed
synchrotron radiation comes from more compact emitting regions). This
dichotomy has been indeed found in the {\it Planck} ERCSC
\cite{Planck11k,gio11}: almost all radio sources show very flat
spectral indices at LFI frequencies, $\alpha_{LFI}\ga-0.2$; at HFI
frequencies BL\,Lacs keep flat spectra ($\alpha_{HFI}\ga-0.5$) while
most of FSRQs show steeper spectra,
i.e. $\alpha_{HFI}<-0.5$. Moreover, the same model gives a remarkably
good fit to all the observed data on number counts and spectral index
distributions of ERS at frequencies above 5 and up to 220\,GHz.

For the above reasons, hereafter we adopt the number counts provided
by the ``C2Ex'' case in \cite{tuc11}.

\subsection{Number counts of ERS in polarization at mm wavelengths}
\label{s61}

To assess the contamination due to undetected ERS in CMB polarization
maps it is necessary to know how many sources can be found with
polarized flux density $P=\sqrt{Q^2+U^2}$\,\footnote{Typically, a
  de--biased estimator is however used for the polarized flux density
  of point sources (as, e.g., $P=\sqrt{Q^2+U^2-s^2}$, where $s$ is the
  uncertainty on $P$; see \cite{saj11} for more details).}  above a
given flux limit $P_{lim}$.  Answering this question involves the
determination of source counts in polarization.  This estimate is
quite difficult to perform directly, by the statistical analysis of
observational data, since the polarized signal is weak and many
samples are usually defined by their completeness in terms only of
total intensity. However, this problem can be overcome by using the
source number counts in total intensity complemented by information on
the statistical properties of the fractional polarization $\Pi$,
expressed in term of some probability function ${\mathcal P}(\Pi)$
\cite{tuc04}.

Let us discuss this point in more detail. Polarization number counts
$n(P)$ can be written as:
\begin{equation}
n(P)=N\,\int_{S_0=P}^\infty {\mathcal P}(P,\,S)\,dS=
N\,\int_{S_0=P}^\infty {\mathcal P}(\Pi,\,S)\,{dS \over S}\,,
\label{ee1}
\end{equation}
where $N$ is the total number of sources with $S\ge S_0$ in the
sample, ${\mathcal P}(P,\,S)$ and ${\mathcal P}(\Pi,\,S)$ are the
probability functions of observing in a source of flux density $S$ a
polarized intensity $P$ and a fractional polarization $\Pi$,
respectively. Assuming that $\Pi$ is independent of $S$, $n(P)$ can
be determined by
\begin{equation}
n(P)=\int_{S_0=P}^\infty {\mathcal P}(\Pi=P/S)\,n(S)
\,{dS \over S}\,.
\label{ee2}
\end{equation}
The probability function ${\mathcal P}(\Pi)$ can be constrained from
the observed distributions of the fractional polarization. In
agreement with \cite{bat11} we model ${\mathcal P}(\Pi)$ by a
log--normal distribution:
\begin{equation}
{\mathcal P}(\Pi)={1 \over \sqrt{2\pi\sigma^2}\Pi}
\exp\Bigg\{-{[\log(\Pi/\Pi_{med})]^2/2\sigma^2}\Bigg\}\,,
\label{ee3}
\end{equation}
where $\Pi_{med}$ is the median of the distribution and
$\sigma^2=1/2\log(\langle \Pi^2\rangle/\Pi^2_{med})$. These formulas
are strictly valid only if $0\le\Pi<\infty$. However, because of the
very low fractional polarization observed in ERS, the upper limit of
$\Pi=0.75$ can be effectively assumed as infinite. In
Fig.\,\ref{f2a} we compared log--normal functions with the
polarization distributions observed in the AT20G survey, confirming
the very good fit of the model with data.

A critical point for our estimates is the variation with frequency of
the fractional polarization observed in ERS.  The data on ERS
polarization discussed in this paper clearly show an higher fractional
polarization at 10--20\,GHz than at few GHz in both flat-- and, more
prominently, in steep--spectrum sources. At higher frequencies, data
become scarce, but there are still indication of a possible further
increase of the polarization fraction from 10--20\,GHz to 40--90\,GHz
\cite{agu10,bat11}. For flat--spectrum sources, this increase could be
due to the combination of two different effects: 1) the polarization
degree actually increases with the frequency; and 2) BL\,Lacs, which
are observed to be more polarized than quasars \cite{agu10}, become
more relevant in number at higher frequencies.

In the following predictions we take into account both effects. We
consider two possible values for the median and the dispersion
$\langle \Pi^2\rangle^{1/2}$ of the log--normal distribution in order
to provide a range of estimates for number counts and power spectra
that could take into account uncertainties in observational data at
$\nu>10\,$GHz. Table\,\ref{t4} reports our estimates for a more
``optimistic'' (lower) and a more ``conservative'' (upper)
case.\footnote{Please note that the choice of the two adjectives,
  ``optimistic'' and ``conservative'', has been done under the point
  of view that polarized ERS do ``contaminate'' CMB maps in
  polarization and, thus, have to be removed from them.} The frequency
dependence of the fractional polarization in ERS is simply modeled by
means of two different sets of median and r.m.s. values at frequencies
below and above about 40\,GHz. Although somewhat arbitrary, this
choice is motivated by the fact that this is the highest frequency at
which multi-frequency polarization samples of ERS are available.

\begin{figure*}
\centering
\includegraphics[width=7.cm]{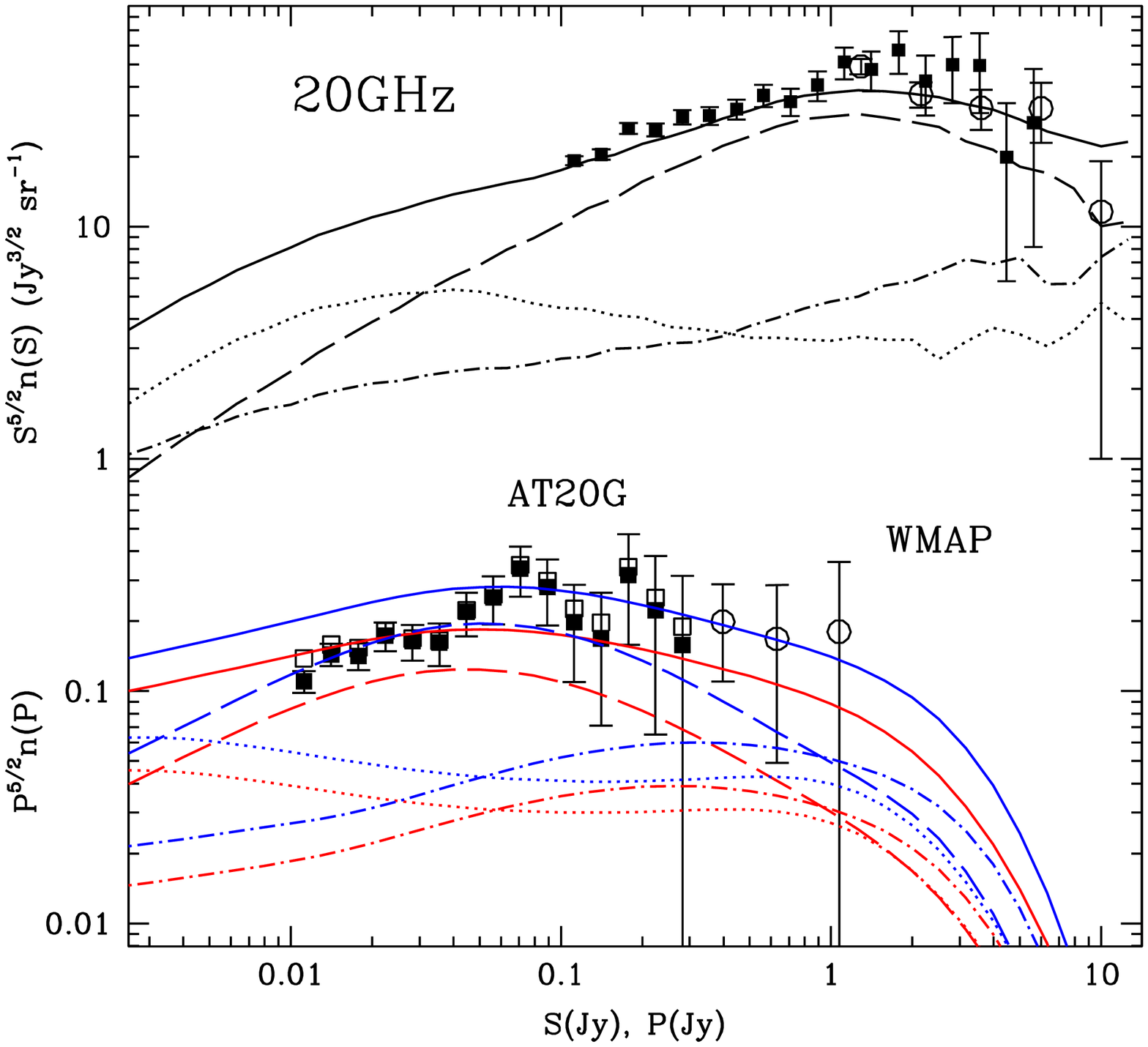}
\includegraphics[width=7.cm]{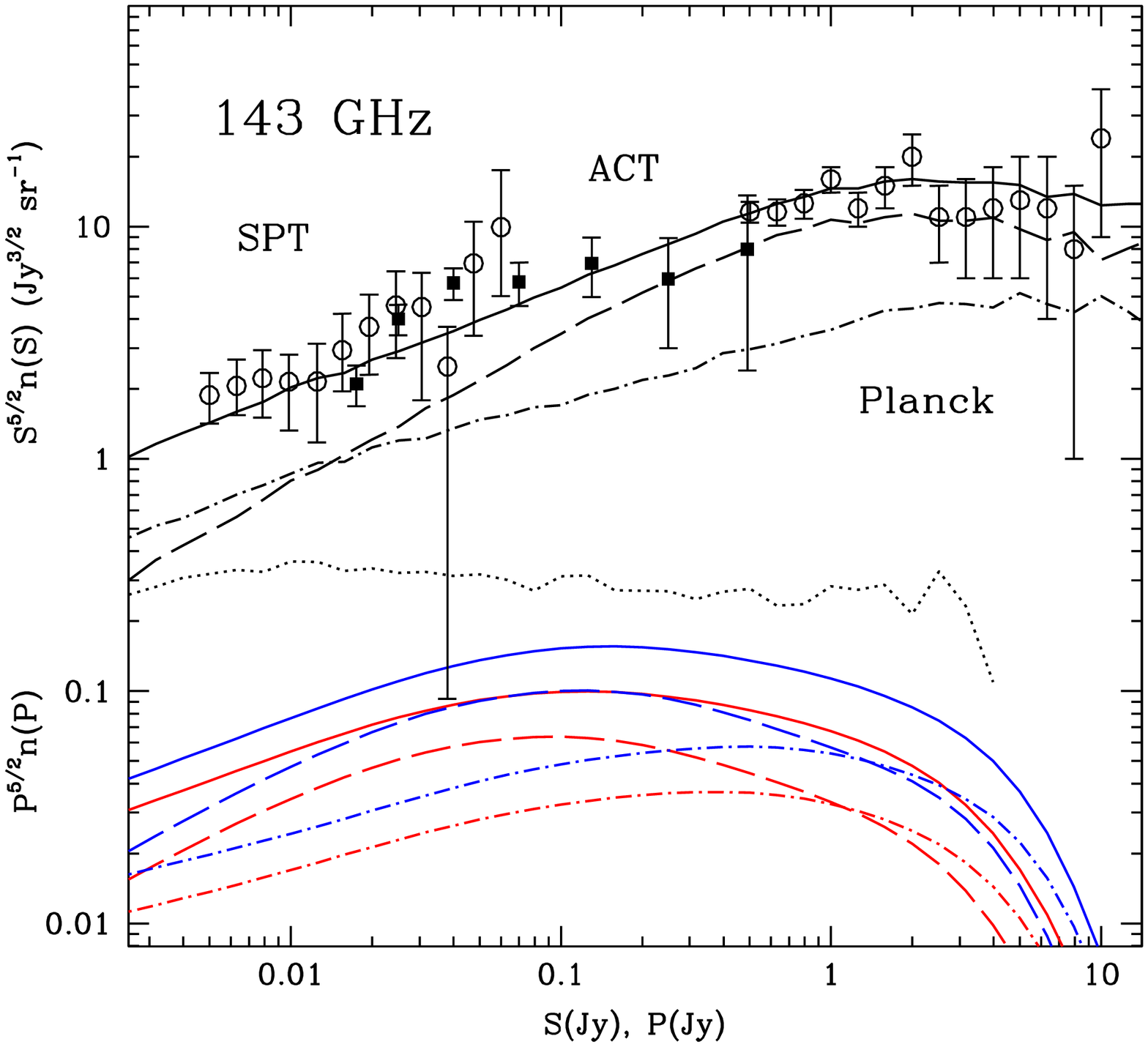}
\caption{Normalized number counts at 20 ({\it left panel}) and at
  143\,GHz ({\it right panel}) in total intensity (upper curves and
  data points) and in polarized intensity (lower curves and data
  points; blue curves are for the more ``conservative'' case and red
  curves for the more ``optimistic'' case). Number counts are for the
  different source populations discussed in the text: solid lines
  represent total number counts; dotted lines are for steep--spectrum
  sources; long--dashed lines are for FSRQs; finally, dot--dashed
  lines are for BL\,Lacs. In the {\it left panel}, filled squares
  represents our current estimates from the AT20G sub-sample (empty
  squares include the contribution from sources without polarization
  detection; see Sect.\,\ref{s5}) and empty circles are from WMAP
  5--years data \cite{lop09}. Data points in the {\it right panel} are
  from published data \cite{vie10,mar11,Planck11i}.}
\label{f5}
\end{figure*}

\begin{itemize}

\item At $\nu\la40\,$GHz we choose a lower and higher median and
  r.m.s. values of the fractional polarization based on the results
  displayed in Fig.\,\ref{f4}. Moreover, we require that number
  counts computed using values in Table\,\ref{t4} be (at least
  partially) compatible with AT20G/WMAP counts in polarization. As
  shown in Fig.\,\ref{f5}, the more ``optimistic'' case fits quite
  well observational counts, especially at $P\la50\,$mJy. On the other
  hand, the more ``conservative'' case tends to overestimate number
  counts at lower polarized fluxes, whereas it fits particularly well
  current data at high $P$ fluxes. This is not unexpected because, in
  this latter case, we take as median and dispersion of the fractional
  polarization for blazars the values provided by \cite{agu10} at
  90\,GHz.

\item At $\nu>40\,$GHz, we make the assumption of an increase of the
  median fractional polarization of about 20\% with respect to the
  corresponding cases at $\nu\la$40\,GHz. This choice is not firmly
  constrained by current data sets; however, we have been guided by
  the fractional polarization levels of ERS observed by \cite{agu10}
  at 86 GHz which fall in the middle between our present lower and
  upper cases.

\end{itemize}

\begin{table}
\caption{Median and dispersion of the log--normal distribution for
the fractional polarization, $\Pi $, as a function of frequency and
for the different radio source populations relevant at CMB
frequencies. Two cases are considered, a more ``optimistic'' one and
a more ``conservative'' one (see sub-Sect.\,\ref{s61}).}
\centering
\begin{tabular}{ccccccc}
\hline
 & \multicolumn{3}{c}{$\nu\la40\,$GHz} &
\multicolumn{3}{c}{$\nu>40\,$GHz} \\
\hline
 & Steep & FSRQ & BL\,Lac & Steep & FSRQ & BL\,Lac \\
\hline
 \multicolumn{7}{c}{lower case (more ``optimistic'')} \\
\hline
$\Pi_{med}$\,(\%) & 3.0 & 2.0 & 2.5 & 4.0 & 2.5 & 3.0 \\
$\langle \Pi^2\rangle^{1/2}$\,(\%) & 5.0 & 3.0 & 3.5 & 6.0 & 3.5 &
4.2 \\
\hline
 \multicolumn{7}{c}{upper case (more ``conservative'')} \\
\hline
$\Pi_{med}$\,(\%) & 4.0 & 3.0 & 3.6 & 5.0 & 3.6 & 4.3 \\
$\langle \Pi^2\rangle^{1/2}$\,(\%) & 6.0 & 3.8 & 4.5 & 7.0 & 4.6 &
5.5 \\
\hline
\end{tabular}
\label{t4}
\end{table}

In Fig.\,\ref{f5} we compare the predicted number counts, $n(P)$, of
ERS in polarized intensity at 20 and 143\,GHz. By comparing the two
panels of the same Figure we can appreciate that $n(P)$ decreases by a
factor $\simeq 1.5$ at 143\,GHz at $P\sim 0.1\,$Jy, in comparison with
the estimated $n(P)$ at 20 GHz. At still fainter polarized
intensities, $P\sim 0.01$\,Jy, this decrement increases due to the
fact that the contribution of steep--spectrum ERS, which are on
average more polarized, becomes less and less relevant at high CMB
frequencies. Finally, the right panel of Fig.\,\ref{f5} indicates a
possible higher relative contribution to total polarized counts,
$n(P)$, coming from BL\,Lacs with respect to FSRQs at 143\,GHz. This
is a direct consequence of the present analysis, which is based on the
observations discussed by \cite{agu10}.

The good agreement with observational data at 20\,GHz, as well as the
reliability of our predictions on ERS number counts in flux density,
$S$, give us confidence in making extrapolations of integral number
counts, $N(>P)$, in total linear polarization at {\it Planck}
frequencies (see Table\,\ref{t5}). It is expected that {\it Planck}
LFI will be able to detect sources in polarization down to $\simeq
200$\,mJy at 30\,GHz and $\simeq 300$--400\,mJy at 44 and 70\,GHz,
whereas {\it Planck} HFI, thanks to a better resolution and
sensitivity, should reach polarized flux limits of
$P\approx100\,$mJy. In both cases, however, a catalogue of only a few
tens of compact ERS should be provided by {\it Planck} in
polarization.

The above quoted detection limits in total polarization are estimated
by the performances foreseen for the {\it Planck} LFI and HFI
instruments and are based on pre--launch measurements
\cite{lea10,ros10} of the detectors calibration and
capabilities. These detection limits take also into account the future
application to the {\it Planck} data -- corresponding to the end of
the nominal mission, i.e. in January 2012 -- of new detection
techniques, specifically designed for detecting compact polarized
sources in CMB maps (see, e.g., \cite{her12} for a recent discussion
on the subject). These techniques have already been applied with
success to WMAP 5-yr maps \cite{lop09}, improving on the results
published by the WMAP team on the same data set.

\begin{table*}
\caption{Expected total numbers of ERS with polarized intensity $\ge
  P_{lim}$ over the full sky at {\it Planck} frequencies. For each
  {\it Planck} channel, the minimum and maximum values here indicated
  refer to our predictions calculated by the more ``optimistic'' and
  by the more ``conservative'' cases previously discussed (see
  Table\,\ref{t4}).}  
\centering
\begin{tabular}{ccccccccc}
\hline $P_{lim}$ & & \multicolumn{7}{c}{$\nu$\,[GHz]} \\
\hline
[mJy]
& & 30 & 44 & 70 & 100 & 143 & 217 & 353 \\
\hline
50 & & 107--164 & 91--140 & 98--151 & 83--129 & 71--109 & 59--89 & 47--70
\\
80 & & 49--77 & 42--66 & 47--74 & 41--64 & 35--55 & 30--46 & 25--37
\\
100 & & 34--53 & 29--46 & 33--52 & {\bf 28--45} & {\bf 25--39} & {\bf
  21--33} & {\bf 18--27} \\
200 & & {\bf 10--16} & 9--14 & 10--16 & 9--15 & 8--13 & 7--12 & 6--10
\\
300 & & 5--7 & {\bf 4--7} & {\bf 5--8} & 4--7 & 4--7 & 4--6 & 3--5 \\
400 & & 3--4 & 2--4 & 3--5 & 3--4 & 2--4 & 2--4 & 2--3 \\
\hline
WMAP$^{(1)}$ & & 8 & 6 & 4 & & & & \\
\hline
\multicolumn{9}{l}{\footnotesize $^{(1)}$ Number of
  polarized ERS detected in the WMAP 5--yr data at $|b|>5^{\circ}$
  by \cite{lop09}}
\\
\multicolumn{9}{l}{\footnotesize
at 33, 41 and 61\,GHz
  ($P_{lim}\approx300\,$mJy).}
\end{tabular}
\label{t5}
\end{table*}

\section{Predictions on the contribution of ERS to the CMB E-- and
  B--modes}
\label{s7}

As for temperature fluctuations, the analysis of CMB polarization
measurements is usually made by the estimate of angular power spectra,
i.e. E-- and B--mode spectra \cite{kam97,zal97}. B--mode polarization,
that arises only from tensor perturbations, is expected to be
extremely weak, and even for the most optimistic cases foreseen by
inflationary models the rms signal is only a fraction of $\mu$K, less
than 1 per cent of the level of temperature anisotropies at degree
scales. On the other hand, polarization of foregrounds (and in
particular of extragalactic sources) is equally shared between E-- and
B--modes \cite{sel97}. ERS are therefore expected to dominate the sky
B--mode polarization at sub--degree angular scales at frequencies
$\nu\la100\,$GHz \cite{tuc05}.


By using the statistical characterization of the fractional
polarization described in the previous Sections, we are able to
estimate the polarized angular power spectra given by undetected ERS
in CMB anisotropy maps. First of all, we assume that ERS follow a
Poisson distribution in the sky. The contribution of clustered ERS to
the angular power spectrum of CMB temperature anisotropy is in fact
small and can be neglected, if ERS are not subtracted down to faint
flux limits (the signal due to clustered ERS becomes more relevant
only at relatively low fluxes, i.e.  $S<10$\,mJy, \cite{gon05,tof05}).

It is well known that an ensemble of Poisson distributed point
sources gives rise to a flat power spectrum of temperature
fluctuations \cite{teg96}. For a sample of sources with flux density
below some cut--off $S_c$, the amplitude of this white noise
spectrum is given by
\begin{equation}
C_{T\ell}=\bigg({dB \over dT}\bigg)^{-2}\,N\langle
S^2\rangle=\bigg({dB \over dT}\bigg)^{-2}\int_0^{S_c}n(S)S^2\,dS\,,
\label{e1}
\end{equation}
where $N$ and $n(S)$ are, respectively, the total and the differential
number of sources per steradian, and $dB/dT$ is the conversion factor
from brightness to temperature,
i.e. $dB/dT\approx10^{-2}$\,$\mu$K/(Jy\,sr$^{-1})\,(e^x-1)^2/(x^4e^x)$
and $x=\nu/56.8$\,GHz.

As an analogy to the expression of the angular power spectrum in total
intensity or CMB temperature, it is possible to define the angular
power spectrum for the Stokes parameters $Q$ and $U$. Because point
sources contribute, on average, equally to $Q$, $U$ and to the
$E$--, $B$--mode power spectra, we assume $C_{E\ell}\simeq
C_{B\ell}\simeq C_{Q\ell}\simeq C_{U\ell}$ (we generally refer to
them as polarization spectra). Following the treatment given by
\cite{tuc04}, we have that
\begin{eqnarray}
C_{Q\ell} & = & \bigg({dB \over dT}\bigg)^{-2}\,N\langle Q^2\rangle =
\bigg({dB \over dT}\bigg)^{-2}\,N\langle S^2\Pi^2\cos^2(2\phi)\rangle
\nonumber \\
& = & \bigg({dB \over dT}\bigg)^{-2}\,N\langle S^2\rangle\langle\Pi^2
\rangle\langle\cos^2(2\phi)\rangle \nonumber \\
& = & 1/2\,\bigg({dB \over dT}\bigg)^{-2}\,\langle\Pi^2\rangle\,C_{T\ell}\,,
\label{e2}
\end{eqnarray}
where the Stokes parameter $Q$ is written in terms of $\Pi$ and of the
polarization angle in the chosen reference system, $\phi$. The factor
1/2 arises because of the uniform distribution of polarization angles.
It is easy to demonstrate that the cross--correlation
temperature--polarization spectra are null, e.g. $C_{TQ\ell}=\langle
S^2\Pi\cos(2\phi)\rangle=0$.

In Eq.\,\ref{ee2} and \ref{e2} we have assumed that the fractional
polarization is independent of the total intensity of the
source. Observations for flat--spectrum sources in the flux density
range $S\ga100$\,mJy seem to support this assumption (see
sub-Sect.\,\ref{s41}). We expect that this is maintained also at
fainter fluxes but only if FSRQs and BL\,Lacs are separated into two
different populations. On the other hand, this hypothesis may not be
true for steep--spectrum sources: a clear anticorrelation between
$\Pi$ and $S$ is observed from data at 1.4\,GHz, whereas, at higher
frequencies, the lack of large samples of steep--spectrum sources does
not allow to determine it. In any case, steep--spectrum ERS are giving
a negligible contribution to number counts at $\nu\geq 100$ GHz and,
thus, this lack of information does not affect our current
predictions.

\begin{figure*}
\centering
\includegraphics[width=6.5cm]{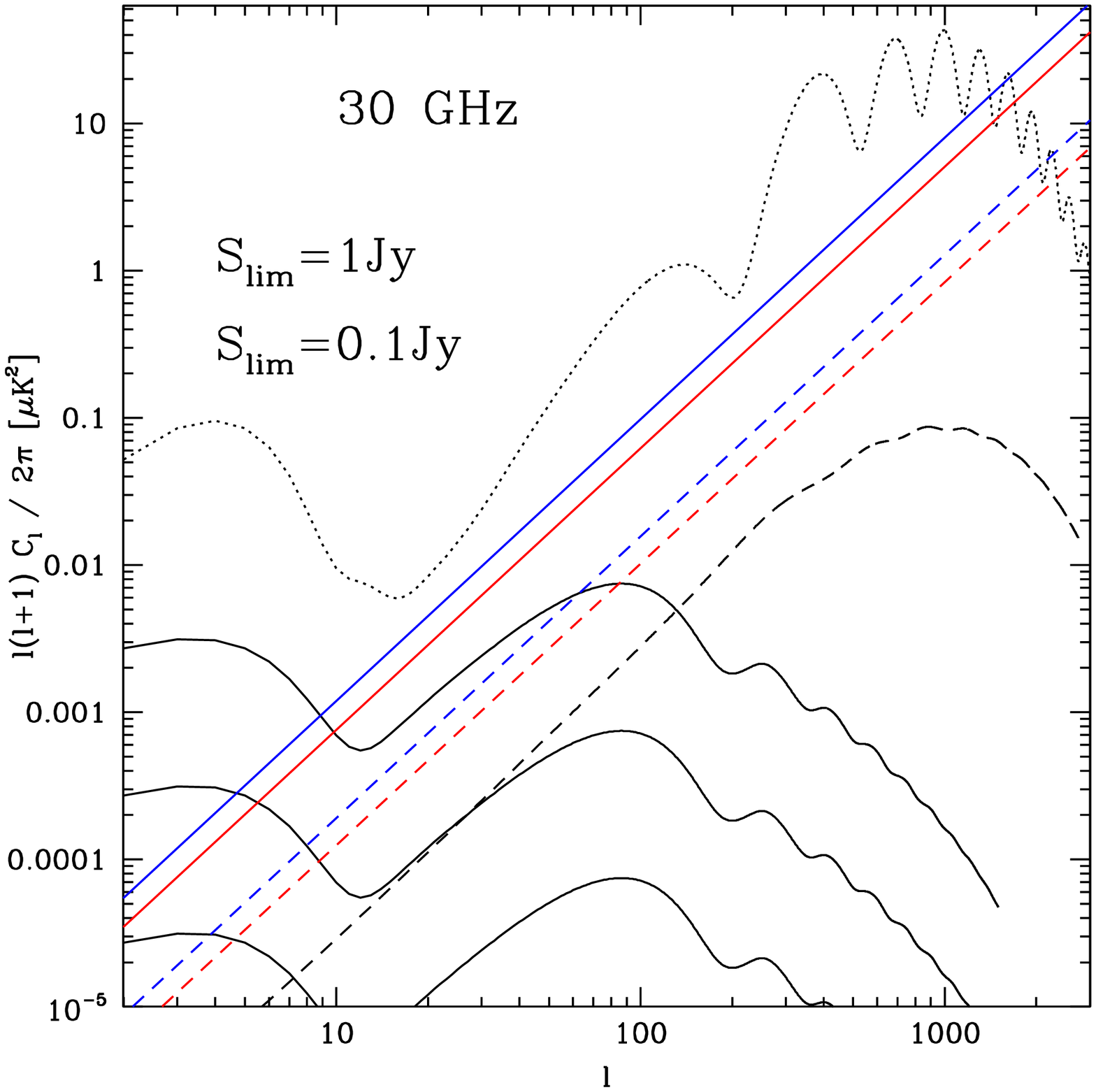}
\includegraphics[width=6.5cm]{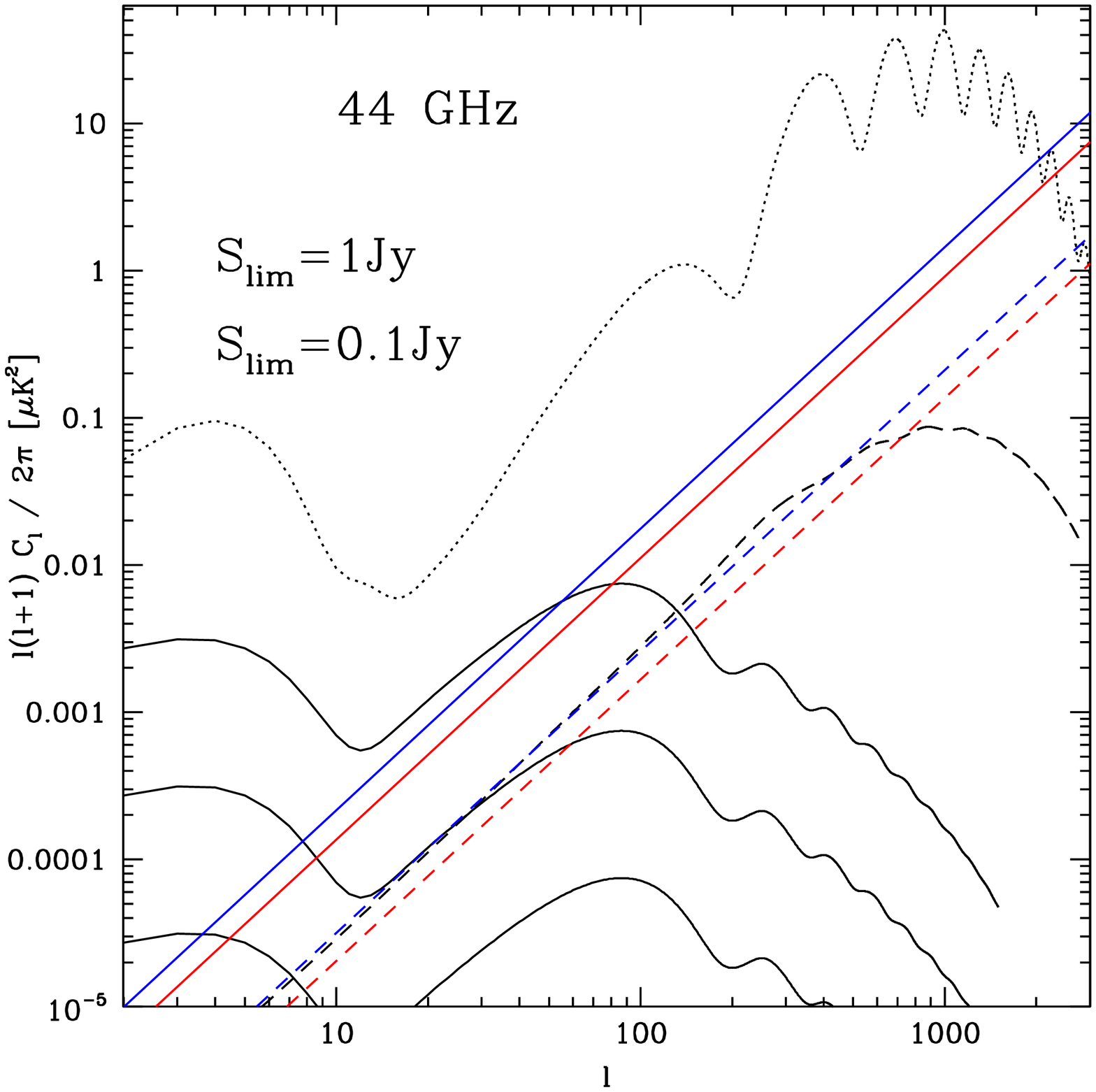}
\includegraphics[width=6.5cm]{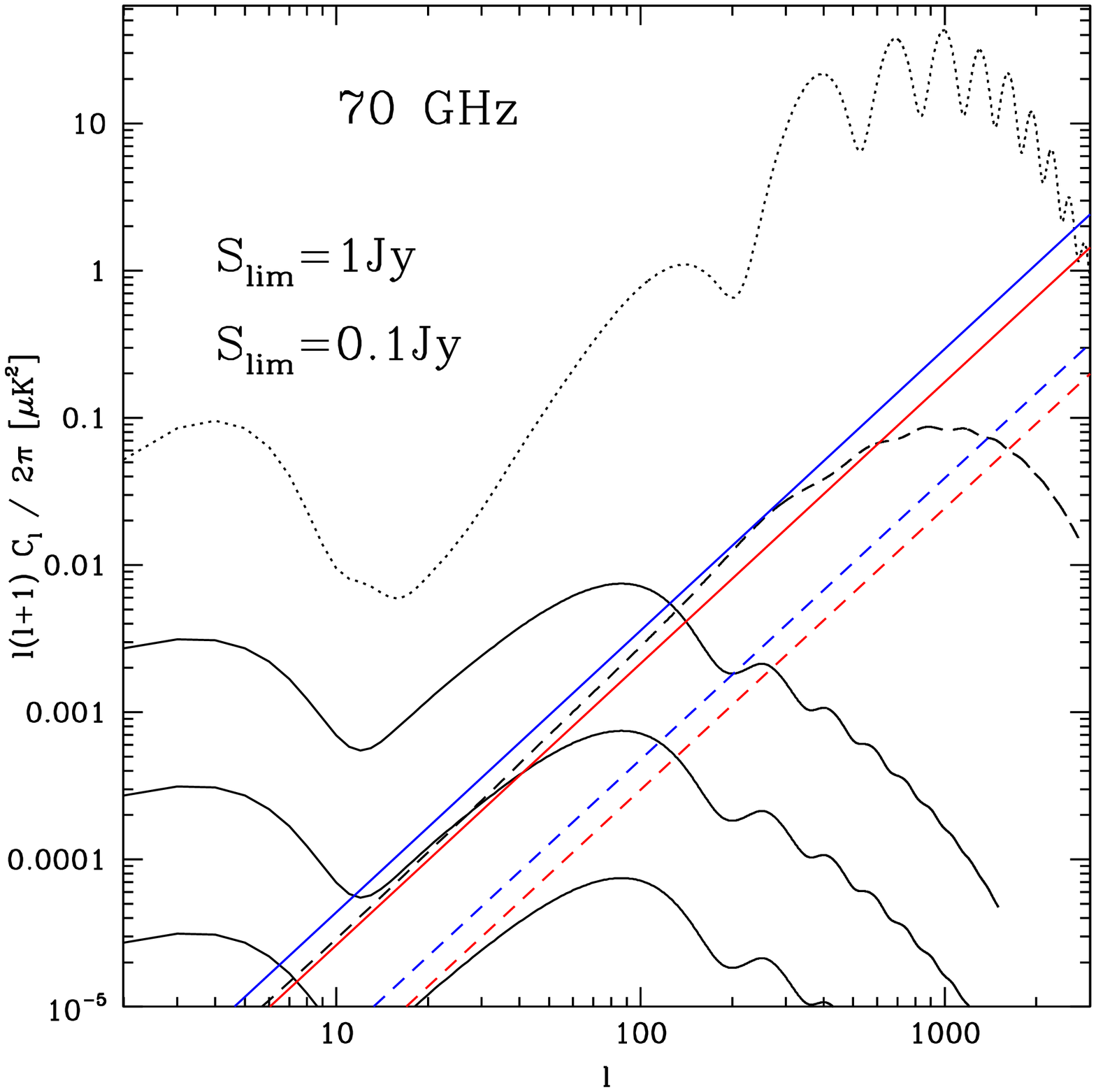}
\includegraphics[width=6.5cm]{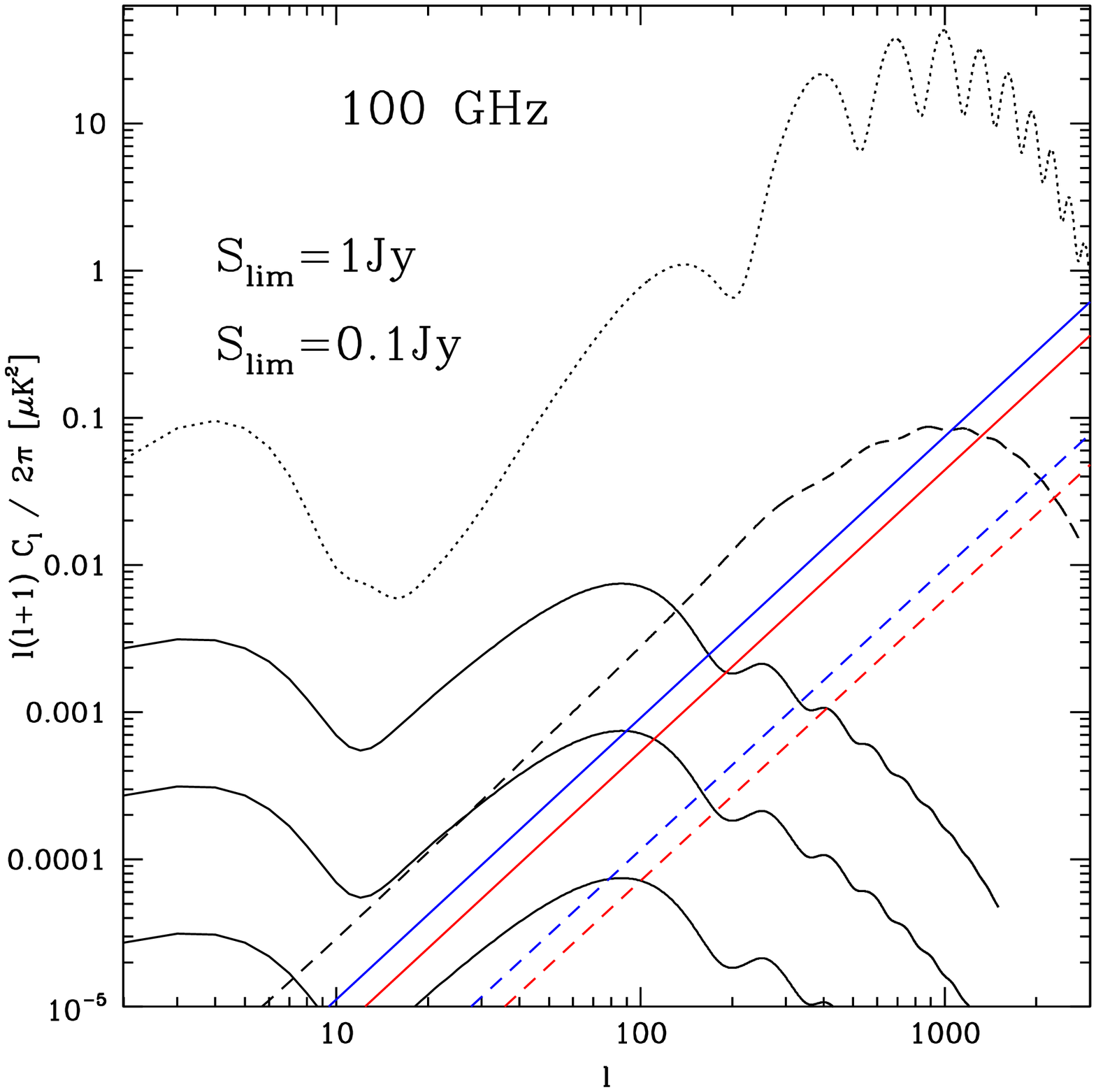}
\includegraphics[width=6.5cm]{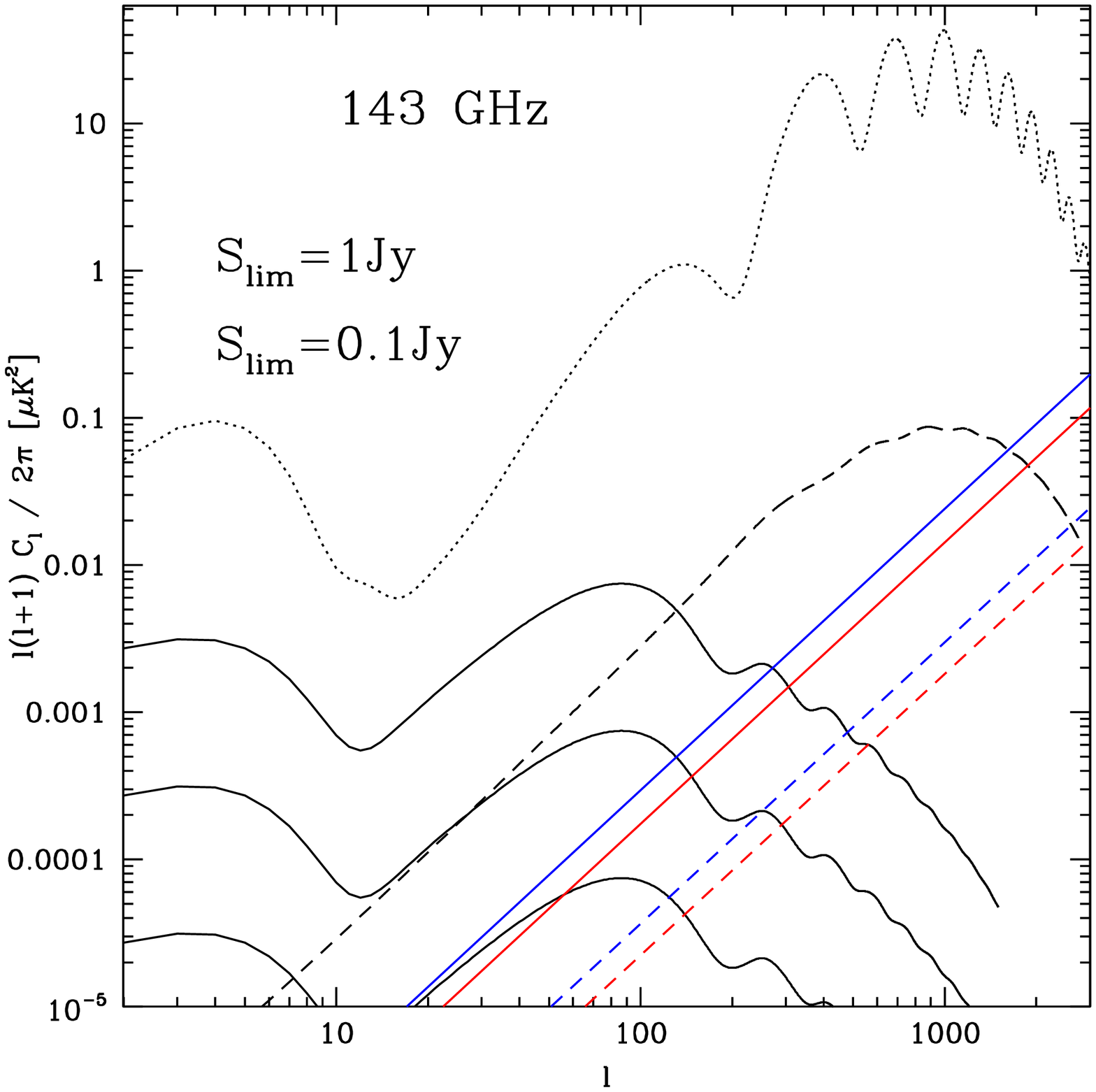}
\includegraphics[width=6.5cm]{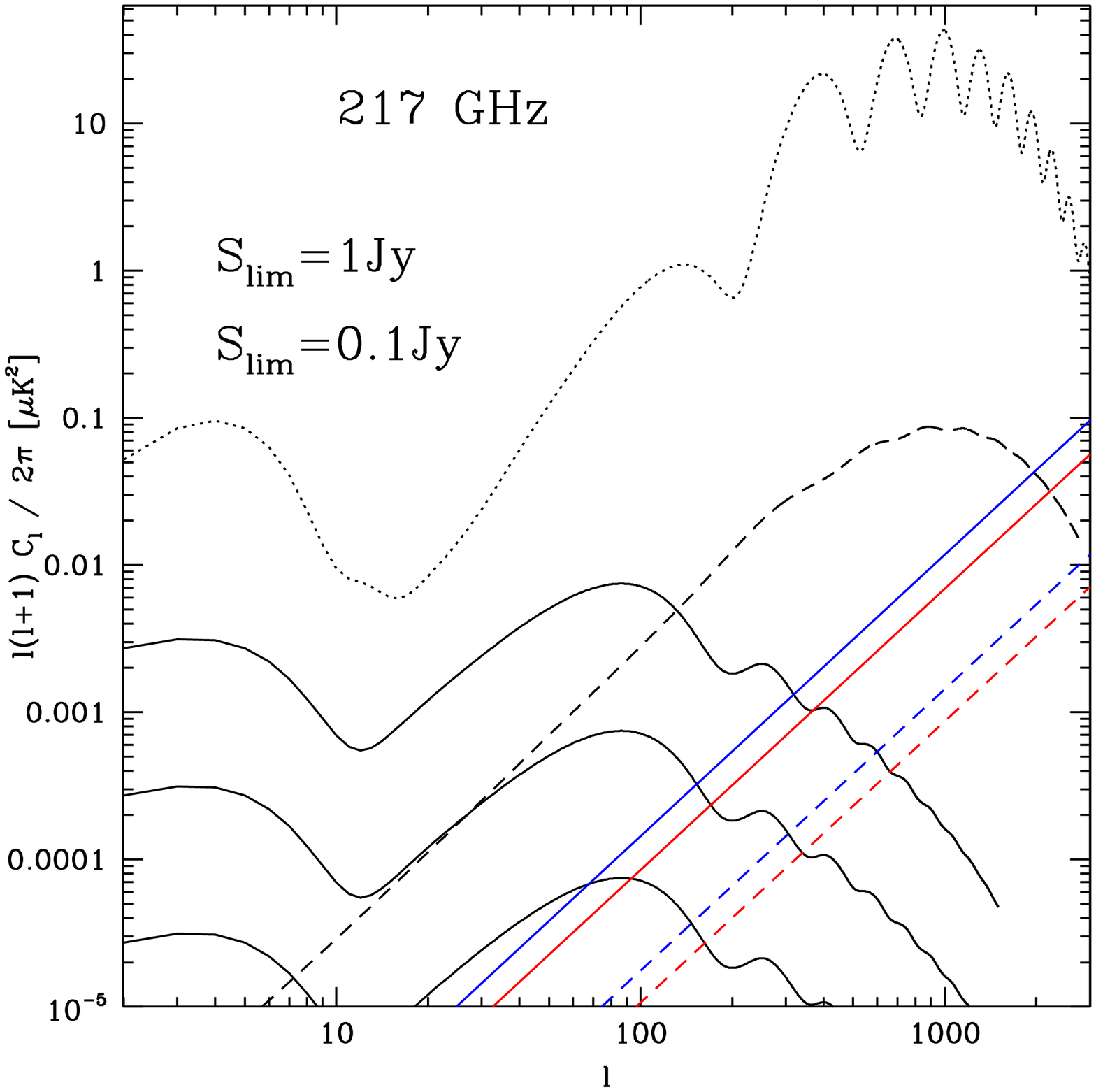}
\caption{Polarization power spectra at six {\it Planck} frequencies
  for the CMB radiation (black lines: dotted lines for the E--mode;
  solid lines for the gravitational wave B--mode with $r=0.1$, 0.01
  and 0.001; dashed lines for the lensing--induced B--mode) and for
  ERS (solid lines is for $S_c=1\,$Jy and dashed lines for
  $S_c=0.1\,$Jy; blue lines correspond to the more ``conservative'' case and
  red lines to the more ``optimistic'' case).}
\label{f7}
\end{figure*}

In Fig.\,\ref{f7} we present the results on ERS polarization power
spectra for the six {\it Planck} frequencies where ERS are
relevant. The value of $\langle\Pi^2\rangle^{1/2}$ is taken according
to Table\,\ref{t4} for the different radio source
populations. Moreover, we consider two cut--offs in flux density:
$S_c=1,$ and 0.1\,Jy. The former value is close to the completeness
limit obtained by the {\it Planck} ERSCS for the frequency channels
$\nu\le100\,$GHz \cite{Planck11i}; the latter one can be seen as a
(somewhat optimistic) reference value for the {\it Planck} high
frequency channels, or a reference value for future experiments.


In Fig.\,\ref{f7} we also plot the CMB power spectrum for the E--mode
and for the B--mode with a tensor--to--scalar ratio $r=0.1$, 0.01 and
0.001. In this way we can have an indication of the level of ERS
subtraction required to allow the detection of a gravitational
wave induced primordial CMB B--mode signal. ERS should not be a strong
constraint on detecting E--mode polarization or the B--modes with
$r\ga0.01$. On the other hand, a primordial CMB B--mode signal
corresponding to lower $r$ values, requires the subtraction of ERS
down to flux detection limits of $\sim100\,$mJy, which will not be
easy, or even possible, with the {\it Planck} sensitivity.

\section{Conclusions}
\label{s8}

In this contribution we have reviewed recently available polarimetric
surveys of extragalactic radio sources at frequencies
$\nu\ga1$\,GHz. These data point out that the typical intrinsic
fractional polarization of ERS is around 2--5\,\% of the total flux
density, $S$, of the source even at frequencies as high as 20\,GHz,
and that in very few objects the fractional polarization is
$\Pi\ga10\%$. This may be due to the low degree of uniformity of
magnetic fields in the internal part of AGN jets and in lobes. Faraday
depolarization is probably the cause of the large number of sources
with a very low level of polarization, i.e. $\Pi\la1\%$, at GHz
frequencies, which also explains the strong increase of the fractional
polarization observed at $\nu\ga10$\,GHz in those objects. This
conclusion is supported by high or extreme values of rotation measure
RM$\gg1000$\,rad\,m$^{-2}$ observed in some blazars (e.g.,
\cite{zav03,bat11}).

Moreover, we have studied how polarization properties of ERS change
from cm to mm wavelengths. For flat--spectrum sources a weak but
constant increase of fractional polarization is observed, with median
(mean) values varying from 1.5\% (2--2.5\%) at 1.4\,GHz, to 2--2.5\%
(2.5-3\%) at 5--10\,GHz and 2--3\% (3--3.5\%) at
10--40\,GHz. Indications that fractional polarization in blazars could
further increase above 40\,GHz come from the recent works by
\cite{bat11,agu10}. On the other hand, a significantly higher
fractional polarization is typically found in steep--spectrum sources,
especially at high frequencies (median is 4--5\% at 10--20\,GHz and
mean between 5 and 6.5\%). However, because of incompleteness of the
samples and of the small number of steep--spectrum sources in surveys
at $\nu\ga10\,$GHz, current observations could be biased by
high--polarized objects.

In general, we do not find any dependence of the fractional
polarization of ERS with the flux density at high radio
frequencies. However, more conclusive evidences require larger and
deeper surveys. Nevertheless, an anticorrelation between $\Pi$ and $S$
in blazars is expected when very faint flux densities are considered
(see Sect.\,\ref{s4}). For the flux--density ranges covered by available
surveys, flat--spectrum sources are dominated by quasars
(FSRQs). These objects are typically less polarized than BL\,Lacs
\cite{agu10}, which become increasingly relevant at fainter flux
densities and become the dominant population at $S\la10$\,mJy
\cite{dez05,pad07}.

We also discuss a formalism to estimate ERS number counts in
polarization and to predict the contribution of unresolved ERS to
angular power spectra at CMB frequencies. As a first application, we
attempt to predict how many polarized ERS the {\it Planck} Satellite
will be able to detect in the different channels sensible to
polarization measurements: we expect that only a dozen polarized ERS
could be detected by {\it Planck} LFI, and a few tens at the HFI
frequencies. Although the number of {\it Planck} detected sources is
low, these data will allow us to study the frequency dependence of the
fractional polarization in a wide range of frequencies, from 30 to
353\,GHz, thus providing original and valuable information on
polarization properties of ERS in the innermost regions of AGN jets.

Finally, our results on polarization power spectra demonstrate that
ERS should not be a strong contaminant to the CMB E--mode polarization
when observing at frequencies $\nu\ga70\,$GHz. Moreover, it seems
unlikely that ERS will have a significant impact on our ability to
detect the B--mode polarization from primordial graviational waves if
$r>0.01$. On the contrary, if the cosmological B--mode signal is
fainter, some strategy will be required to subtract the confusion
noise produced by radio sources with flux--density $S<1\,$Jy. At
sub--mm wavelengths, where radio sources become less and less
relevant, confusion noise of point sources may be dominated by dusty
galaxies (e.g., \cite{neg07}). Polarization of these objects should be
relatively low, but they are expected to be significantly
clustered. Although the level of this effect is still quite uncertain,
polarization spectra from dusty galaxies could begin to dominate the
one produced by ERS already at $\nu\ga 200$-300\,GHz, at least at
angular scales relevant for the detection of the B--modes.

\section*{Acknowledgement}

The authors thank the referee, R.B. Partridge, for his insightful
comments and criticisms that helped a lot in clarifying the main
assumptions of the paper and also in improving its final
presentation. MT acknowledges financial support from the French
``Centre National d'\,\'{E}tudes Spatiales'' (CNES). LT acknowledges
partial financial support from the Spanish Ministry of Science and
Innovation (MICINN), under project AYA2010--21766--C03-01.

\end{document}